\documentclass[]{aa}  
\usepackage{hyperref}
\usepackage{graphicx}
\usepackage{pdflscape}
\usepackage[utf8]{inputenc}
\usepackage[T1]{fontenc}
\usepackage{placeins}
\usepackage{txfonts}

\usepackage{lscape}
\usepackage{lineno}
\begin{document}

   \title{S-Process Nucleosynthesis in Chemically Peculiar Binaries}


   \author{A.J. Dimoff \inst{1}\fnmsep\inst{2}
          \and
          C.J. Hansen \inst{2} 
          \and 
          R. Stancliffe \inst{3}
          \and
          B. Kub\'atov\'a \inst{4}
          \and
          I. Stateva \inst{5}
          \and
          A. Ku\v cinskas \inst{6}
          \and
          V. Dobrovolskas \inst{6}
          }

   \institute{Max Planck Institute f\"ur Astronomie,
              K\"onigstuhl 17, 69117 Heidelberg Germany 
              \email{dimoff@mpia.de}
         \and
             G\"othe Universit\"at Frankfurt Institute of Advanced Studies,
             Ruth-Moufang-Straße 1,60438 Frankfurt am Main Germany
        \and
            H.H. Wills Physics Laboratory, University of Bristol, 
            Tyndall Avenue, Bristol BS8 1TL, UK 
        \and
            Astronomical Institute, Academy of Sciences of the Czech Republic,
            Fri\v{c}ova 298, 251 65 Ond\v{r}ejov, Czech Republic 
        \and 
            Institute of Astronomy and National Astronomical Observatory, Bulgarian Academy of Sciences,
            Boulevard "Tsarigradsko shose" 72, 1784 7-Mi Kilometar, Sofia, Bulgaria 
        \and 
            Institute of Theoretical Physics and Astronomy, Vilnius University, 
            Saul\.{e}tekio al. 3, Vilnius 10257, Lithuania 
             }

   \date{Accepted 23. September 2024}

 
  \abstract
   {Around half of the heavy elements in the universe are formed through the slow neutron capture (s-) process, which takes place in thermally pulsing asymptotic giant branch (AGB) stars with masses $1-6\;M_{\odot}$. The nucleosynthetic imprint of the s-process can be studied by observing the material on the surface of binary barium (Ba), carbon (C), CH, and carbon-enhanced metal-poor (CEMP) stars.}
   {We study the s-process by observing the luminous components of binary systems polluted by a previous AGB companion. Our radial velocity (RV) monitoring program establishes an ongoing collection of binary stars exhibiting enrichment in s-process material for the study of elemental abundances, production of s-process material, and binary mass transfer.}
   {From high resolution optical spectra, we measure radial velocities (RVs) for $350$ stars and derive stellar parameters for $\sim150$ stars using ATHOS. For a sub-sample of $24$ chemically interesting stars we refine our atmospheric parameters using ionization and excitation balance with the Xiru program. We use the MOOG code to compute one-dimensional local thermodynamic equilibrium (1D-LTE) abundances of carbon, magnesium, s-process elements (Sr, Y, Zr, Mo, Ba, La, Ce, Nd, Pb), and Eu to investigate neutron capture events and stellar chemical composition. We estimate dynamical stellar masses via orbital optimization using Markov chain Monte Carlo techniques in the ELC program, and we compare our results with low-mass AGB models in the FUll-Network Repository of Updated Isotopic Tables \& Yields (FRUITY) database.}
   {In our abundance sub-sample, we find enhancements in s-process material in spectroscopic binaries, a signature of AGB mass transfer. We add the element Mo to the abundance patterns, and for 12 stars we add Pb detections or upper limits, as these are not known in the literature. Computed abundances are in general agreement with the literature. Comparing our abundances to dilution-modified FRUITY yields, we find correlations in s-process enrichment and AGB mass, supported by dynamical modeling from RVs.}
   {From our high-resolution observations, we expand heavy element abundance patterns and highlight binarity in our chemically interesting systems. We find trends in s-process element enhancement from AGB stars, and agreements in theoretical and dynamically modelled masses. We investigate evolutionary stages for a small sub-set of our stars.} 
   
   \keywords{nuclear astrophysics -- s-process -- binaries: spectroscopic, radial velocities -- stars: chemically peculiar, abundances}

   \maketitle
%
\section{Introduction}\label{sec:INTRO}
Nucleosynthetic s-process events occur in low-mass thermally-pulsing asymptotic giant branch (AGB) stars \citep{1957RvMP...29..547B}, within the He inter-shell region. Protons are ingested into the helium burning zone, converting $^{12}\mathrm{C}$ to $^{13}\mathrm{N}$, which then decays to $^{13}\mathrm{C}$, providing a strong neutron source via the $^{13}\mathrm{C}(\alpha,\mathrm{n})$ reaction. The excess neutrons produced through this channel lead to a succession of n-captures and $\beta$-decays. These captures and decays cause the production of heavy elements up to Pb. 

The s-process material synthesized in the helium inter-shell region is brought to the stellar surface by Third Dredge-Up events (TDUs) and violent convective motion in the envelope \citep{1998ApJ...497..388G}. This heavy metal enriched material is expelled from the AGB star by strong stellar winds during thermal pulses, and can be accreted onto a binary companion.

The s-process signature in AGB stars is characterized by comparing the abundances of the first (N $\sim$ 50, including Sr, Y, Zr) and second s-process peaks (N $\sim$ 82, including Ba, La, Ce) \citep{2001ApJ...557..802B, 2018A&A...620A.146C}. Stars of different mass will produce different patterns of elements, owing to their different interior properties and AGB lifetimes. 

Observationally there are two ways of learning about s-process nucleosynthesis. If one observes AGB stars (intrinsic S-stars) \citep{2018A&A...620A.148S, 2021A&A...650A.118S} directly, the signature of ongoing s-process nucleosynthesis can be seen in the stellar atmosphere in the presence of heavy elements such as Sr, Ba, Tc, and Pb. The detection of Tc in the stellar spectrum indicates ongoing s-process nucleosynthesis and is the most robust method to identify intrinsic S-stars; other machine learning methods have been recently proposed \citep{2019AJ....158...22C}. For reviews on AGB stars, see \citet{2005ARA&A..43..435H, 2006NuPhA.777..311S, 2014PASA...31...30K}, and \citet{2022Univ....8..220V}. More evolved AGB stars display high carbon-to-oxygen ratios (C/O > 1) and are known as carbon stars \citep{2023EPJA...59...17S}.

One can also observe extrinsic systems where heavy elements produced by an AGB star have been transferred onto a less evolved binary companion, which shows radial velocity variation \citep{1999A&A...345..127V}. The binaries enriched by an AGB star generally fall into two categories depending on their metallicity and carbon enrichment: the metal-rich barium (Ba) and CH stars \citep{1980ApJ...238L..35M, 1983ApJ...268..264M}, and the carbon-enhanced metal-poor -s (CEMP(-s)) stars. In these systems, the observed star has received s-process material from a former AGB companion, which has since become a faint white dwarf.

At higher metallicities [Fe/H] $\gtrsim -1$, the AGB mass transfer in binary systems can be followed by studying Ba or CH stars \citep{2018A&A...620A.146C, 2021MNRAS.505.5554S}. Despite having known about Ba stars for more than 50 years \citep{1951ApJ...114..473B}, they are perhaps the least well studied of the s-process stars in terms of their element patterns. Recent works have significantly improved the efforts of studying these stars from the nucleosynthetic perspective: \citet{2016MNRAS.459.4299D} and \citet{2019IAUS..343...89C} studied 5 elements (Y, Zr, La, Ce, Nd) in 182 and 169 Ba giants respectively, and \citet{2021MNRAS.507.1956R} provides knowledge about a handful of elements (Sr, Nb, Mo, Ru, La, Sm, and Eu) in 180 Ba giants; but this is not enough to fully identify the patterns of elements produced by AGB stars.

At metallicities [Fe/H] $< -2$ we trace the s-process in the early Galaxy through the CEMP-s stars. The CEMP-s stars show strong molecular C-N bands, are typically very old ($>$ 10 Gyr), and are important to understand the detailed composition of the early s-process \citep{2019A&A...623A.128H}. The vast majority of these stars are indeed binaries (\citet{2014MNRAS.441.1217S, 2016A&A...588A...3H, 2018A&A...620A..63A}), and many reside in wide binaries with long orbital periods up to thousands of days. To better understand the physics of the s-process, we perform a comprehensive study of the heavy elements produced by AGB stars. 

Monitoring radial velocities over a long baseline in time \citep{2016A&A...588A...3H} identifies the binarity of Ba, CH, CEMP-s stars, and other chemically peculiar systems that may form through similar processes. To date, most approaches have either been made in small samples ($< 15$ stars) with a long baseline, or larger samples over a short period of time. We improve on previous approaches by computing abundances for 11 heavy elements and monitoring radial velocities in a large sample of $350$ stars over the long baseline of four years. Orbital parameters of binary star systems inform us about the masses of stars involved and about the way mass is transferred; a crucial aspect in binary stellar evolution modelling. Comparing masses derived from chemical abundance patterns with masses derived from orbital parameters, we have two independent methods to constrain the donor AGB star mass. Access to the telescopes within the Chemical Elements as Tracers of the Evolution of the Cosmos - Infrastructures for Trans-National Access (ChETEC-INFRA TNA)\footnote{\url{https://www.chetec-infra.eu/ta/}} network has been critical in our investigation of nuclear astrophysics.

This paper is divided into sections. Sample selection, data acquisition and reduction, and analysis procedures are detailed in Section \ref{sec:DATA}. Outcomes and results in Section \ref{sec:RESULTS}. A discussion on the implications of these results follows with Section \ref{sec:DISCUSSION} with a summary of our findings in Section \ref{sec:CONCLUSION}. 

\section{Data}\label{sec:DATA}

\subsection{Sample selection}\label{sec:SAMPLE}

We source our targets from catalogs of cool stars, supplemented by querying large surveys. We select focused catalogs by stellar classifications of interest: AGB, Ba, CEMP-s, C, and CH type stars, including both intrinsic and extrinsic S stars. Our source focused catalogs include \citet{1984PW&SO...3....1S, 2001BaltA..10....1A, 2020A&A...639A..24E, 2016ApJ...833...20Y, 2019MNRAS.483.3196C, 2021A&A...654A.140K}, and \citet{2018A&A...620A.146C}. We supplement our focused list with targets selected from the Apache Point Observatory Galactic Evolution Experiment (APOGEE), \emph{Gaia}, GALactic Archaeology with HERMES (GALAH), and Large Sky Area Multi-Object Fiber Spectroscopic Telescope (LAMOST) surveys, where we query the databases on metallicity, heavy element enrichment, and binarity indicators (details below). 

The APOGEE survey samples major populations of the Milky Way with moderate resolution ($R\sim22500$) and high signal-to-noise ($>100$) infrared ($1.5 - 1.7 \mu m$) spectra \citep{2017AJ....154...94M}. The catalog includes stellar parameter estimation, metallicities, and sparse heavy element abundances. We query the \emph{Gaia} DR3 \citep{2023A&A...674A...1G} database for targets with more than $10$ transits, radial velocity uncertainties $>$ 5km/s, and astrometric reduced unit weight error (\texttt{RUWE}) $\gtrsim$ 1.4 to increase the chances of selecting binaries. \emph{Gaia} DR3 includes medium resolution RVS spectra ($R\sim11000$) in the near-IR around the calcium triplet, providing selection criteria based on indicators of enrichment in s-process material, including Ce and Nd. We also select based on output from secondary data products like the The Final Luminosity Mass Age Estimator \citep[FLAME,][]{FLAME:LL:OLC-036}, a CU8/Apsis software, including estimates of the masses, ages, and orbital parameters of stars in \emph{Gaia} DR3. The GALAH survey provides a wealth of spectroscopic data for bright stars in the southern hemisphere \citep{2019A&A...624A..19B} in relatively high resolution ($R\sim28000$). Chemical compositions and orbital properties for this sample are available for reference and comparison. GALAH allows selection targets identified as Ba-enhanced, CEMP-s, or other C-enhanced stars, and we select candidates in both confirmed and suspected binary systems. We select stars from the LAMOST catalog \citep{2012RAA....12.1197C} showing heavy element features in their spectra (for example, Ba and Sr), are optically bright enough for our network of telescopes, and display binary star characteristics. 

\begin{figure}[ht]
    \centering
    \includegraphics[width=\linewidth]{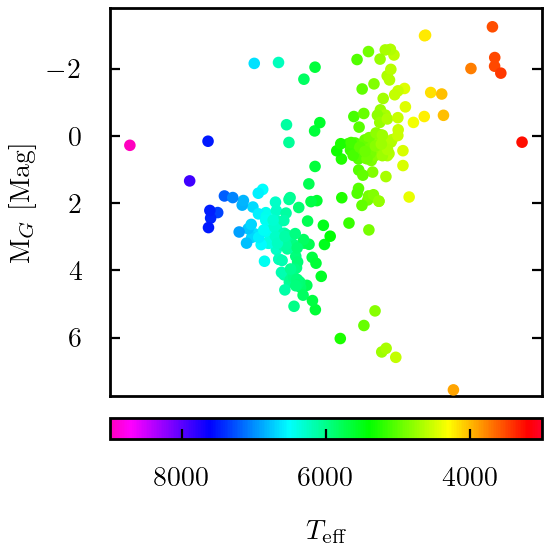} 
    \caption{Color-magnitude diagram for a representative group from our sample using \emph{Gaia} photometric data and parallax. The color gradient follows temperature along the x-axis, with warmer objects in blue on the left and cooler objects in red on the right. Our full sample is generally split between main-sequence dwarfs and giants, with temperatures ranging from 3500 - 8000 K.}
    \label{fig:gaia_HRD}
\end{figure}

From these catalogs and surveys, we compile a sample of $350$ targets (limited to stars brighter than $\sim$13th magnitude), displayed on a Hertzsprung-Russell (HR) diagram in Figure \ref{fig:gaia_HRD}. Our observational sample is a combination of warm to cool (spectral types (A),F,G,K) dwarfs that exhibit s-process enrichment, and giant stars either known to produce or exhibit the presence of s-process material (AGB, Ba, CEMP, C, CH). Many of our targets are either in known binaries, or suspected to be in binary systems based on \emph{Gaia} DR3 parameters \texttt{RUWE} and \texttt{radial\_velocity\_error}. This large sample of stars constructs our radial velocity database, each to be observed $6-8$ times over the full range of $\sim4$ years, depending on the number of previous RV data points. A subset of these stars are selected for high signal-to-noise ratio (S/N) targeted observations to investigate their surface chemical composition.

\subsection{Observations}\label{sec:OBSERVATIONS}
Our observing strategy is effectively split in two: we obtain high S/N ($\gtrsim50$ at around $4500$ \AA$\;$) spectra to derive abundances of heavy elements (within $\pm$0.2 dex) from stars with peculiar abundances or in known binaries, and we collect snapshot spectra of S/N $\gtrsim 15-20$ suitable for precise RV measurements and long-term monitoring of stars with peculiar abundances that may reside in binary systems. 

\begin{figure}[]
    \centering
    \includegraphics[width=\linewidth]{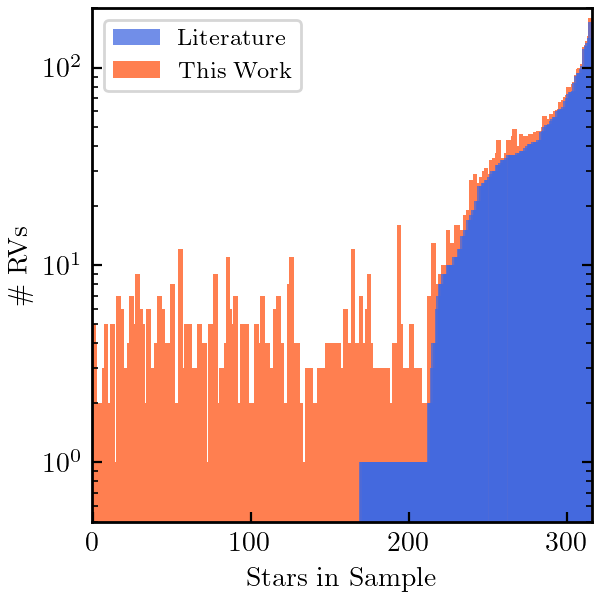}
    \caption{Contributions from RV monitoring program to the literature. Each bin is one observed star, and the y-axis value is the total number of RV data points for the given star. Blue data is the available literature data, and orange is the contribution from our monitoring program.}
    \label{fig:RV_contributions}
\end{figure}

We make use of high-resolution (R $\gtrsim$ 30000 up to 67000) echelle spectrographs, available through the Trans-National Access (TNA) as part of the ChETEC-INFRA framework, and through the Max Planck Institute for Astronomy. These instruments allow high precision RV measurements and precise abundance calculations. Each of our observation facilities is briefly described, with Table \ref{tab:obs_nights} summarizing the progress of the ongoing observing program, and Figure \ref{fig:RV_contributions} displays our contributions to the RV literature. The RV monitoring effort will be continued through the lifetime of the ChETEC-INFRA network in 2025. We perform RV monitoring from all five of our observatories. The MPIA instruments are better equipped for high-S/N observations with larger telescope mirrors; our abundance sub-sample mainly comes from these instruments, and we include the highest quality spectra from the TNA observatories. 

\begin{table}[]
    \centering
    \caption{Awarded nights of observation for each instrument between 2021 and 2024.}
    \begin{tabular}{l c c r}
        \hline \hline
        Instrument & R & Telescope Size & Total Nights \\
        \hline
        VUES   & 37000 & 1.65m & 39 \\
        OES    & 40000 & 2.00m & 28 \\
        ESpeRo & 30000 & 2.00m & 21 \\
        FIES   & 67000 & 2.65m & 10 \\
        FEROS  & 48000 & 2.20m & 35 \\
        \hline
    \end{tabular}
    \tablefoot{This list also includes nights lost due to bad weather conditions, Target of Opportunity (ToO) programmes, or other reasons.}
    \label{tab:obs_nights}
\end{table}

\paragraph{Vilnius University Moletai Astronomical Observatory (MAO)}
The MAO at Vilnius University in Lithuania hosts a 1.65m Ritchey-Chretien telescope, with the fibre-fed Vilnius University Echelle Spectrograph (VUES) at the Cassegrain focus \citep[][]{2014SPIE.9147E..7FJ, 2016JAI.....550003J}. With a resolution of $R=37000$ and a wavelength range of $4000 - 8800$ \AA, VUES is an excellent instrument to measure RVs with high precision. Estimated velocity uncertainty with the VUES instrument is on the order of $0.8$ km/s. This instrument is mostly used for RV monitoring, and the highest S/N spectra are of abundance measurement quality.

\paragraph{Astronomical Institute of the Czech Academy of Sciences (ASU) Ond\v{r}ejov Astronomical Observatory}
The Ond\v{r}ejov Observatory is part of the ASU, which operates the 2m Perek telescope on which the Ond\v{r}ejov Echelle Spectrograph (OES) is mounted \citep[][]{2004PAICz..92...37K}. The spectrograph is fed by a thorium-argon calibration lamp and a flat field calibration lamp, and has a spectral resolution power $\lambda / \Delta\lambda \approx 40000$ in the H$\alpha$ region (6562 \AA) and a spectral coverage between 3753-9195 \AA. This instrument is well-suited for RV measurements, and spectra of high S/N can be used for chemical abundance estimations. 

\paragraph{Bulgarian National Astronomical Observatory Rozhen} 
The Echelle Spectrograph Rozhen (ESpeRo) \citep{2017BlgAJ..26...67B} is a cross-dispersed fibre-fed instrument obtaining spectra from 3900 \AA$\;$to 9000 \AA$\;$at high resolutions from $R\sim30000-45000$. Another ideal instrument for RV measurements, the average RV uncertainty of our observations using ESpeRo is $\sim 0.5\;km/s$. Stacked observations from multiple exposures or long exposure times result in higher S/N, and have been used to compute stellar abundances. 

\paragraph{Roque de los Muchachos Observatory, Instituto Astrofisico de Canarias (IAC)} 
Located on the island of La Palma, in Canarias, Spain in accordance with the  IAC, the 2.65m Nordic Optical Telescope (NOT) is the home of the high-resolution FIbre-fed Echelle Spectrograph (FIES). The instrument is fully described in \citet{2014AN....335...41T}. The FIES instrument is a cross-dispersed high-resolution echelle spectrograph mounted in an independent building for thermal and mechanical stability. We use the highest resolution setting of $R\sim67000$ on the high-resolution 1.3 arcsecond fibre number 4. The optical range is from $3700 - 8300$ \AA, without gaps. Our average RV uncertainty from FIES is on the order of $0.01\;km/s$. We find FIES to be a consistent instrument and useful in our abundance investigation. 

\paragraph{La Silla Observatory, European Southern Observatory (ESO)} 
The Fibre-fed Extended Range Optical Spectrograph (FEROS) \citep{1998SPIE.3355..844K, 1999Msngr..95....8K} is a temperature and humidity controlled echellograph, mounted on the MPG/ESO 2.2m telescope at the La Silla Observatory in Chile. With a resolution of $R\sim48000$, the RV accuracy can be as good as $21\;m/s$ over a 2 month baseline. Our measured average RV uncertainty from this instrument is of this order, $\sim0.02\;km/s$. High-S/N FEROS spectra are suitable to compute stellar abundances. \\

From our compiled sample of observable stars, we have observed $350$ stars with an average of about 4-5 observations per star. We have focused our RV monitoring efforts on stars showing heavy element enhancements, which are expected to be in binaries, yet have few observed RVs in the literature. Our sample includes spectroscopically confirmed barium stars. RV follow up is planned for stars where abundances have only recently been measured. This paper represents the first results of the RV monitoring program, as well as results from the first abundance sub-sample. The full RV and abundance samples will be analyzed and presented in future works. We note that not all of the stars with measured RVs are included in this paper. 

\subsection{Data reduction}\label{sec:REDUCTION}
Because we collected data from multiple sources, data are reduced in different ways corresponding to the instrument, each briefly described here. The raw VUES spectroscopic data are reduced using a spectroscopic reduction pipeline written in IDL. The bias is first subtracted, spectral orders are traced using halogen lamp (flat field) images, a flat correction is applied, and the wavelengths are calibrated using Thorium-Argon lamp spectra. Finally the science spectra are extracted along the traced orders, and the dispersion solution is applied. Data from the VUES instrument is provided to the end user in a pre-reduced format, in 3D \texttt{.fits} files. For a final step in our data reduction, we normalize the spectra by performing a blaze correction using the flat field frame, and fitting the resulting continuum with a polynomial.

Data from OES are reduced using the Ond\v{r}ejov Echelle Spectrograph REDuction (OESRED) semi-automatic reduction pipeline (see \cite{2023Cabezas}), tailored to reduce 2D OES spectra in IRAF. This pipeline includes wavelength and heliocentric calibration and continuum normalization. The biases, halogen lamp flat fields, and thorium-argon frames are combined for their respective master images. The master bias is subtracted from the master flat field and the science frames. The apertures for the spectral orders in the master flat, lamp, and science frames are determined by fitting high-order polynomials. Wavelength calibration with the master lamp is done for each night of observation. Flat correction effectively removes the blaze function from all orders as well as fringing in the reddest orders. Final normalization of the individual orders is done with high-order splines ($\sim10$) in batches.

Raw CCD data from the ESpeRo spectrograph at the Rozhen observatory is reduced by the observing staff following standard spectral reduction procedures in IRAF, including bias subtraction, flat field correction, and wavelength calibration. The master bias is removed from the master flat field and sciences frames. Individual echelle orders are extracted from the master flat, the best obtained stellar spectrum, or a standard (RV or photometric) bright star that has been observed. The orders are extracted for the stellar spectra, master flat, and thorium-argon (ThAr) images. Science orders are divided by the flat field, and wavelengths are calibrated using the ThAr lamps. A heliocentric correction is applied to the spectra. We normalize the spectra via polynomial fitting in IRAF. 

Full spectral reduction for the data collected from FIES and FEROS is performed using the Collection of Elemental Routines for Echelle Spectra (CERES) pipeline \citep{2017PASP..129c4002B}. We use tools and routines within the code for CCD image reduction, tracing of the echelle orders, optimal extraction of the wavelength solution, and RV estimation. The CERES pipeline provides normalized optical spectra from the FEROS and FIES data, and we merge the spectral orders for a complete 1D spectrum by interpolating the overlapping spectral regions. 

Echellogram spectra have a curved `blaze' shape, where they receive more flux in the middle of the chip compared to the edges. To flatten and normalize the spectra from the TNA telescopes, we divide the science apertures by the flat fields, and fit the resulting spectra using splines of high order ($\sim 7-10$) to ensure a continuum value of $\approx 1$. Normalization is required for RV cross-correlation routines, stellar parameter estimation, and accurate abundance determinations. Spectra from the ChETEC-INFRA TNA telescopes can be noisy on edges of the orders, there can be large regions of spectral overlap between orders, and gaps between orders. Because of this, we do not merge orders for these TNA instruments. 

\subsection{Stellar parameters}\label{sec:STARPARM}
Stellar atmospheric parameters are necessary for generating synthetic stellar atmospheres and computing abundances. To homogenize our samples from different instruments, we use ATHOS (A Tool for HOmogenizing Stellar spectra) from \citet{2018AA...619A.134H}, because it is a fast and automated code needed for our sample size. The ATHOS code estimates effective temperature $\mathrm{T_{eff}}$, surface gravity $\log g$, and metallicity [Fe/H] by associating flux ratios (FRs) around the Balmer lines H$\alpha$ and H$\beta$ ($4800$ \AA$\;$and $6500$ \AA), as well as iron features between $5000$ and $5500$ \AA. Spectral normalization is extremely important in measuring flux ratios, and can have a strong effect on the resulting parameter estimation.

The ATHOS program works on stellar spectra with resolution within the range $R \sim 2000-67000$. ATHOS is a machine-learning algorithm trained on chemically normal stars, operating effectively within temperatures $4000-6500$ K, surface gravities from $1-5$, and metallicities between $\sim$-4.5 to $\sim$+0.3 dex, covering spectral types F, G, and K. Uncertainties in the parameters determined by ATHOS stem from statistical uncertainties within the training set. The parameters are estimated in order: first temperature, then surface gravity, and finally metallicity. If the estimate of the temperature is highly uncertain, this will propagate to the estimation of the surface gravity, which further compounds into the estimate of the metallicity. 

ATHOS can quickly determine atmospheric parameters for a large sample of stars, provided their parameters are within that of the training set. Since the program depends on flux ratios from a set of wavelength ranges, it suffers degeneracies when stars are chemically peculiar or have appreciable rotational velocities \citep{2018AA...619A.134H}.

In addition to estimating $\mathrm{T_{eff}}$, $\log g$, and $\mathrm{[Fe/H]}$, we compute the microturbulence parameter $\xi$ using an empirical relation from \citet{2017A&A...604A.129M}:

\begin{equation}
    \xi_t = 0.14 - 0.08\;\mathrm{[Fe/H]} + 4.90 \left( \frac{\mathrm{T_{eff}}}{10^4}\right) - 0.47 \log(g).
\end{equation}

While this formula is designed to work with metal-poor giants, we find it sufficient to estimate the microturbulence of our sample from ATHOS.

Our full observational sample slightly exceeds the parameter limits of the ATHOS program, so we run ATHOS on a trimmed sample of our $350$ observed stars. The stars outside the quoted ranges will be analysed in a future paper. Not all of our spectra are of high enough quality to estimate accurate parameters using ATHOS; we estimate accurate parameters in a high-quality sub-sample of about $150$ stars, which can be seen in Figure \ref{fig:Kiel_Diag}.


The Xiru code \citep{AlencastroPuls_phd} is a program to find spectroscopic parameters of a star ($T_{eff}$, $\log(g)$, [M/H], and microturbulent velocity $\xi$) from excitation/ionisation balance given a set of equivalent widths of Fe I/II spectral lines. ARES (Automatic Routine for Equivalent widths of Spectra) \citep{2015A&A...577A..67S} is a tool for measuring equivalent widths of spectral features. We use separate line lists for metal rich stars \citep{2010A&A...513A..35A, 2016A&A...587A.124K} and metal poor stars \citep{2012A&A...545A..31H, 2016A&A...587A.124K} with ARES to compute equivalent widths of Fe I and II features in our observed stars.

Choosing a subset of our stars analyzed by ATHOS, we refine the stellar parameters by using the Xiru program using Fe I and II equivalent widths measured with ARES. A summary of the parameters for our abundance-quality sub-sample is in Table \ref{tab:atmos_params}, where the first line for each star has our Xiru parameters and the second line has those from the literature, with references in the last column.

Comparing many sets of atmospheric parameters, we must make a choice in which ones we use to generate the comparison model atmosphere and perform the abundance analysis. We perform a spectral fitting comparison for each set of atmospheric parameters in temperature and pressure sensitive spectral regions around H$\beta$, H$\alpha$, and the Mg triplet, where there are also Fe I and Fe II lines, to determine the best fitting spectral model. The parameters estimated with Xiru through ionization and excitation balance consistently provide best fit models to the observed spectra. Comments on atmospheric parameters are included in the discussion in Section \ref{sec:DISCUSSION}.

We use our stellar parameters determined using Xiru to generate synthetic spectra from the ATLAS9 Kurucz stellar atmosphere models \citep{2002A&A...392..619H,2003IAUS..210P.A20C}. We interpolate the model atmosphere grid between nearest points in $\mathrm{T_{eff}}$, $\log(g)$, [Fe/H], and $\xi$ space for custom fit models to our stars.

To summarize our observational efforts, we measure RVs of $350$ stars with 6-8 repeats to determine binary orbital parameters. For these stars, we use ATHOS to estimate atmospheric parameters where possible. To isolate our abundance measurement quality spectra, we make cuts on spectral S/N ($> 40$ around $\lambda\sim4500 \AA$) and ATHOS parameter uncertainties ($\Delta \mathrm{T_{eff}} < 300$ K, $\Delta \log g < 0.5$, $\Delta \mathrm{[Fe/H]} < 0.3$ dex), and use Xiru on the highest quality subset. Within our sub-sample of $24$ selected stars for which we compute stellar abundances, the average uncertainties in $T_{eff}$, $\log g$, and [Fe/H] are $89$ K, $0.40$, and $0.16$ respectively

In addition to high S/N, these $24$ stars satisfy at least one of our criteria of interest: they are chemically interesting stars (Ba, C/CH, CEMP stars), or are flagged as chemically interesting candidates (Sr or Ba candidates from LAMOST for example, or exhibit high Ce enhancements in \emph{Gaia} RVS spectra). They are either in known binaries, or are known to be chemically interesting and in suspected binaries based on \emph{Gaia} RUWE parameter or relatively high radial velocity errors. Some stars in the sample are RS CVn stars, which are active F, G, or K stars known to be in binaries and are worth investigating from the nucleosynthetic perspective.

\subsection{Stellar abundances}
We compute 1D-LTE abundances from spectral features using the MOOG software \citep{2012ascl.soft02009S,2023AAS...24222704S}, with the PyMOOGi\footnote{\url{https://github.com/madamow/pymoogi/}} implementation from \citet[][]{2017AAS...23021607A}. We use the \texttt{synth} driver to compute abundances via synthetic spectra comparison for blends and hyper-fine split lines and the \texttt{abfind} driver for equivalent width fitting in clean atomic lines. A $\chi$-squared routine within \texttt{synth} indicates the best fit synthetic spectrum to the localized observed spectrum. 

Spectral lines of each element are selected from the NIST database\footnote{\url{https://physics.nist.gov/}}, and atomic data (excitation potentials and $\log(gf)$) are taken from \texttt{linemake} \citep{2021RNAAS...5...92P}. We determine the total abundance of an element by averaging the abundances measured from each of the lines. A line list for our atomic species can be found in Table \ref{tab:line_list}, with wavelengths, excitation potentials and oscillator strengths, constructed with data from \texttt{linemake}. NIST data quality flags are included where available. Lines with hyper-fine or isotopic splitting are marked accordingly, with (hfs). 

\subsection{Orbital parameter determination}
We measure RVs from our observed spectra using the cross-correlation method. For the TNA telescopes, we compare the spectra order-by-order to a set of stellar templates of different spectral types that have been manually shifted to the rest frame. Given the large number of orders in the echellograms, this provides a statistically robust method to compute the RV, and uncertainties are estimated using the standard deviation across the spectral orders. For FIES and FEROS, the CERES pipeline performs a cross-correlation to compute RVs across the full spectrum. 

We model the orbits of our program stars using the ELC code. The ELC program is a photodynamical modeling software for binary stars \citep{2000A&A...364..265O}. The code is general, and the orbits of a variety of binary systems can be directly modeled, including eclipsing and RV variable systems. Further details can be found in \citet{2019AJ....157..174O} and \citet{2023AJ....166..114D}. 

We collect available RV data and orbital parameters for our targets from literature sources as well as The RAdial Velocity Experiment \citep[RAVE,][]{2006AJ....132.1645S} and the 9th Catalogue of Spectroscopic Binary Orbits \citep[SB9,][]{2004A&A...424..727P}, including references therein. A representative list of input parameters is presented in Table \ref{tab:orbit_params_out} in the second line for each star. With our RV time-series data and informed priors from the literature, we use ELC to optimize and refine the orbital and physical parameters of the binaries with a differential evolution Markov chain Monte Carlo routine \citep[]{2006S&C....16..239T} to best fit the input data. 

In an eccentric orbit, the RV semi-amplitude of the primary component $K_1$ can be measured from spectroscopic observations \citep{1992PASP..104..270M, 1995Natur.378..355M}, and is expressed as

\begin{equation}
    K_1 = \left( \frac{2\pi G}{P} \right) ^ {1/3} \frac{M_2 \sin i}{(M_1 + M_2)^{2/3}} \frac{1}{(1 - e^2)^{1/2}},
\end{equation}

\noindent where $P$ is the orbital period of the binary, $M_1$ and $M_2$ are the masses of the primary and secondary components, $i$ is the orbital inclination with respect to the observer, and $e$ is the orbital eccentricity. The same expression exists for the secondary component, with the subscripts switched. 

From a single-lined spectroscopic binary with one visible component, the individual masses of the components cannot be measured directly; instead masses are approximated through the mass function via Kepler's third law:

\begin{equation}\label{eqn:mass_func}
    f(m) = \frac{M_2^3}{(M_1 + M_2)^2}\sin ^3 i = 1.0385 \times 10^{-7} K_1^3 (1-e^2)^{3/2} P.
\end{equation}

Both component masses and the inclination are degenerate, and depend on one another to keep the mass function constant for the system. The observable parameters $K_1$, $e$, and $P$ allow computation of the mass function. 

However, the masses of the binary components can be estimated by constraining the inclination of the system and simultaneously solving for both masses. For non-eclipsing systems, we set upper limits on the inclination $i \lesssim 75^{\circ}$, and we estimate a lower limit of $i \gtrsim 20^{\circ}$ because we detect significant RV variations. The inclination is further constrained by assuming rotational synchronization of the stars with their orbit; i.e. they rotate in the plane of the orbit, at the same rate that the stars orbit each other. Since our sample is assumed to be mostly older stars typically with long periods and lower eccentricities, we assume enough time has passed such that the rotational rate of the star has synchronized with the orbit. The parallax angle $\Pi$ and semi-major axis $a$ can be used to compute a relation between the period and masses following \cite{2017A&A...608A.100E}:

\begin{equation}
    \frac{a}{\Pi} = P^{2/3} \frac{M_2}{(M_1+M_2)^{2/3}}.
\end{equation}

With \emph{Gaia} DR3 parallaxes and magnitudes we estimate stellar luminosities and, using the relation $L = 4\pi R^2 T_{eff}^4$, we can estimate the radius of the luminous component. This allows estimation of the luminous mass, which we use as input for the ELC program. Where available, we use mass estimates from the FLAME pipeline and mass information from references within Table \ref{tab:orbit_params_out} as informed priors for our orbital optimization. It is noted that the estimated mass of evolved stars from the FLAME pipeline generally are less accurate and have larger uncertainties. 

From \citet{2019A&A...626A.127J}, the average mass of a barium giant is $\sim2.5\;M_{\odot}$, and varies between about $1.5$ - $3.0\;M_{\odot}$; we initialize our visible components using this mass range in our higher metallicity systems. For low metallicity stars, we initialize with a mass of approximately $1 M_{\odot}$, where the average mass of a CEMP star is about $0.8 M_{\odot}$. If the unseen component in the binary is indeed an AGB remnant, it should be a white dwarf. From AGB stars between 0.85 - 7.5 M$_{\odot}$ initial mass, white dwarf remnant masses are typically $0.5\;M_{\odot} < m < 1.1\;M_{\odot}$ \citep{2018ApJ...860L..17E}, and we use this lower limit in our optimization. We set an upper limit for the mass of white dwarfs with the Chandrasekhar mass $M_* \leq 1.44\;M_{\odot}$ \citep[][]{1931ApJ....74...81C}. 

\section{Results}\label{sec:RESULTS}

\subsection{Atmospheric parameters}
\begin{figure}[t]
    \centering
    \includegraphics[width=\linewidth]{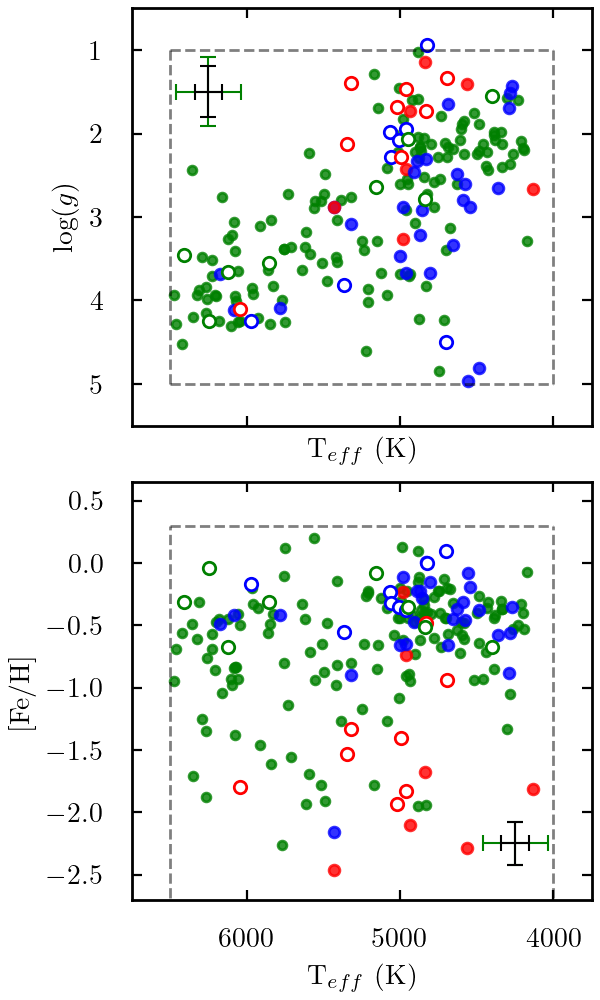}
    \caption{TOP: Kiel diagram of our estimated stellar parameters from ATHOS and Xiru, with ATHOS operational limits as the dashed grey box. Open circles are parameters estimated by Xiru. Blue data points are known Ba stars, red are carbon enhanced stars, and green are ``other'' stars. Floating error bars are for Xiru (black) and ATHOS (green). Cool giants are in the upper right, and warm dwarfs are in the lower left. BOTTOM: Metallicity vs temperature for our this sub-sample, with ATHOS operational limits as the dashed grey box. Colors are the same as the top panel.}
    \label{fig:Kiel_Diag}
\end{figure}

High-resolution spectra allow for good estimation of atmospheric parameters, an important step in determining the abundances of elements on the surface of the star. After making the cuts from our full sample of $350$ stars, we arrive at a refined sample of about $150$ stars. Surface gravities and effective temperatures of our stars analyzed with ATHOS, overlaid with parameters in our abundance sub-sample from Xiru, are displayed in a Kiel diagram in the top panel Figure \ref{fig:Kiel_Diag}, with metallicities in the bottom panel. In Figure \ref{fig:Kiel_Diag}, blue data points are known Ba stars, red data points are C-enhanced stars, and green data points are ``other'' stars, yet unclassified by their abundances or chemical peculiarity. The open circles are the Xiru parameters for our abundance sub-sample. Error bars in the figure are the average uncertainties across our sample.

Typically, barium stars and CH stars have metallicities between $-1.0 \lesssim [Fe/H] \lesssim 0.0$, where about $75\%$ of our sample lies. About $25\%$ of our sample is stars with lower metallicity ([Fe/H] $< -1$), including CEMP-s stars. After trimming our sample to the ATHOS range, the effective temperature range is $4100 - 6500$ K, surface gravities range from $0.65 - 4.97$ dex, and metallicities range is $-2.6 - +0.20$ dex. 

We find Xiru sometimes estimates higher temperatures and metallicities compared to ATHOS and other studies when determining atmospheric parameters. This may be due to forcing abundances of Fe I and Fe II to match; over-ionization of Fe II is a known problem in cool giants, or NLTE effects at lower metallicities. Our sample of ATHOS stellar parameters for $\approx150$ stars is made available on the Milne-Center GitHub page\footnote{\url{https://github.com/Milne-Centre/Barium-Star-Repository/}}. 

\subsection{Abundances}
We investigate signals of enrichment from AGB nucleosynthesis through s-process abundance patterns. Here we present our relative abundances and patterns for our sub-sample of 24 stars. We construct atomic and molecular line lists using \texttt{linemake} including the elements we want to measure, as well as potential molecular contaminants in blended lines.

\begin{figure}[t]
    \centering
    \includegraphics[width=\linewidth]{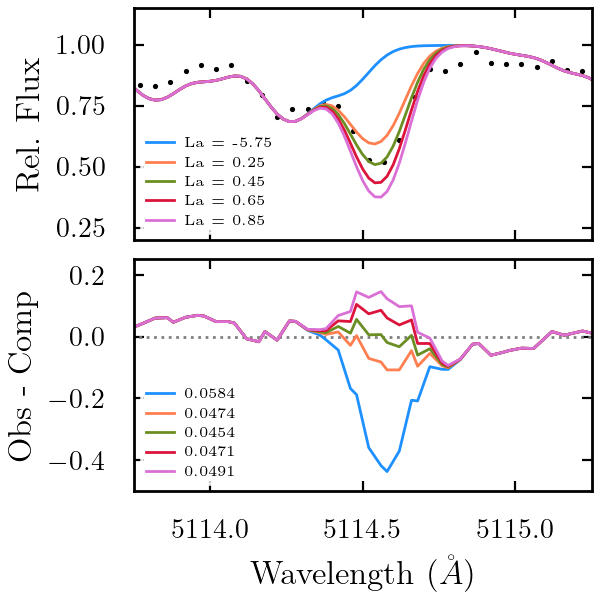}
    \caption{Top Panel: Synthetic fit to a carbon and lanthanum blend at 5114.5\AA\;in the CEMP-s star HE 0414-0343. Bottom Panel: Residuals} in the synthetic spectral fits and the observed data points.
    \label{fig:synth_fit}
\end{figure}

We compute abundances for the elements C, Mg, Fe, Sr, Y, Zr, Mo, Ba, La, Ce, Nd, Eu, and Pb for our abundance sub-sample , or identify upper limits using synthetic spectral fitting. An example of a synthetic spectrum fit for the star HE 0414-0343 can be seen in Figure \ref{fig:synth_fit}, where black data points are the observed spectrum, and multi-colored solid lines are synthetic spectral fits. The blue line is set with absolute abundances of $\log\epsilon = -5$, or effectively no contribution to provide a baseline to the fit. In this blended line, carbon dominates the spectrum, but there is a significant contribution from lanthanum. The carbon abundance is first determined from the 5156 \AA\;Swan band. In the figure, the green line is the best fit for La, with $\log\epsilon_{La} = 0.45$. This corresponds to a [La/Fe] ratio of 1.30, which we find compatible with the heavy element abundances found by \citet{2016A&A...588A..37H} considering differences in atmospheric parameters; see Table \ref{tab:abund_compare}. 

\begin{figure*}[]
    \centering
    \includegraphics[width=\textwidth]{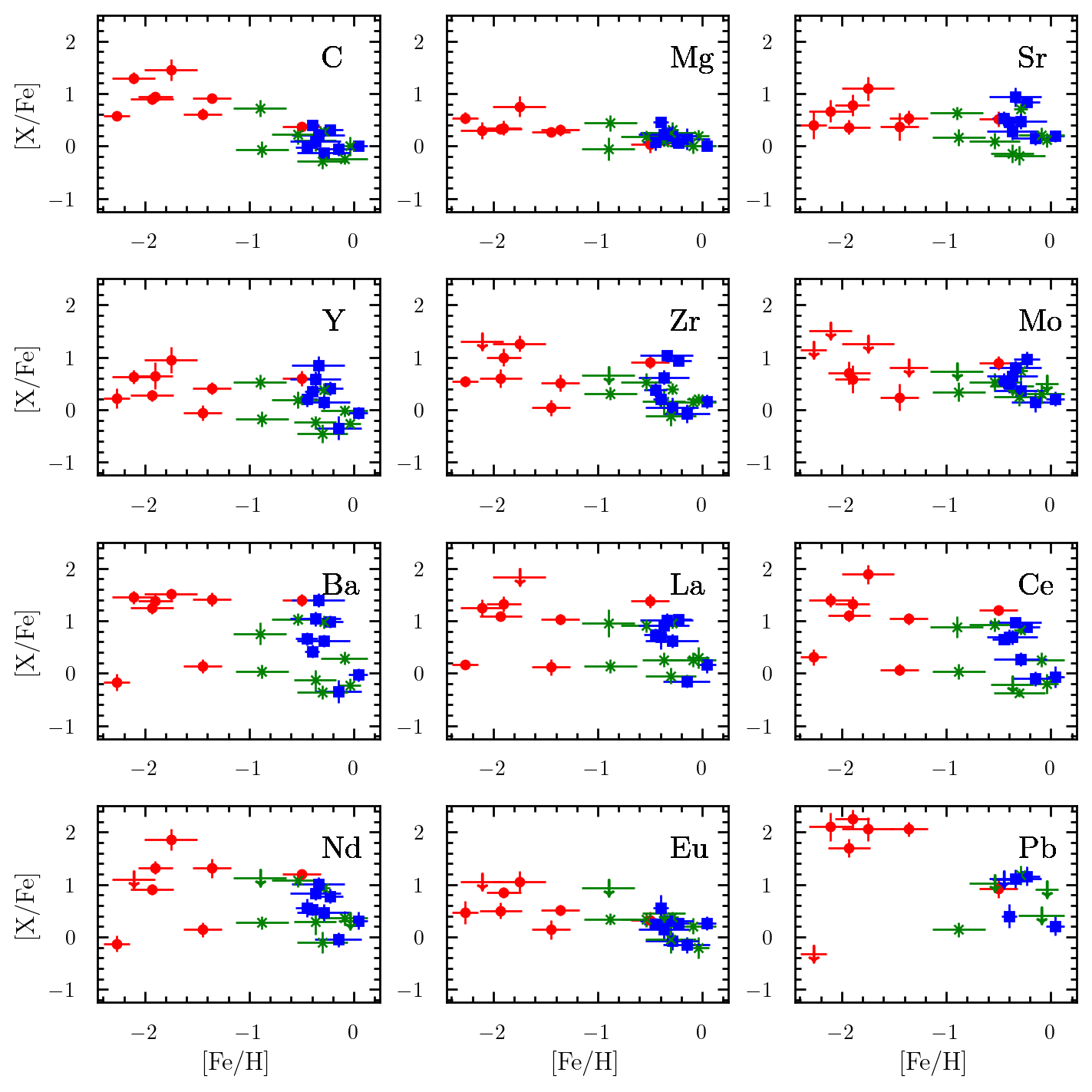}
    \caption{Abundances for our sample of stars by element. Blue squares are Ba stars, red circles are carbon-enhanced (CEMP-s/-no or CH) stars, and green x's are ``other'' or unclassified stars. Inverted arrows are upper limits on the abundance of [X/Fe]}
    \label{fig:abund_grid}
\end{figure*}

Abundances $\mathrm{[X/Fe]}$ for our sub-sample are compared to the literature star-by-star in Table \ref{tab:abund_compare}, where the upper row for each star is our derived abundances, and the lower row is that of the literature, with references. Quoted uncertainties are the average statistical uncertainties between lines of the same element. Abundances are plotted in Figure \ref{fig:abund_grid}, where red circles are carbon-enhanced (CEMP-s/-no, CH) stars, blue squares are known barium stars, and green x's are ``other'' stars, yet unclassified by their heavy metal content or carbon enrichment. 

Across the sub-sample, our computed abundances typically agree with the literature within the combined uncertainties, and larger differences arise from scaling of stellar parameters, mainly metallicity. However, some of our derived abundances do not agree with the literature, particularly at higher metallicities where [Fe/H] $>$ -0.5; this could be due to NLTE effects, differences in model atmosphere parameters temperature or surface gravity, or higher spectral resolution \cite[][]{2014ApJ...781...40P}.

\begin{figure*}[]
    \centering
    \includegraphics[width=\textwidth]{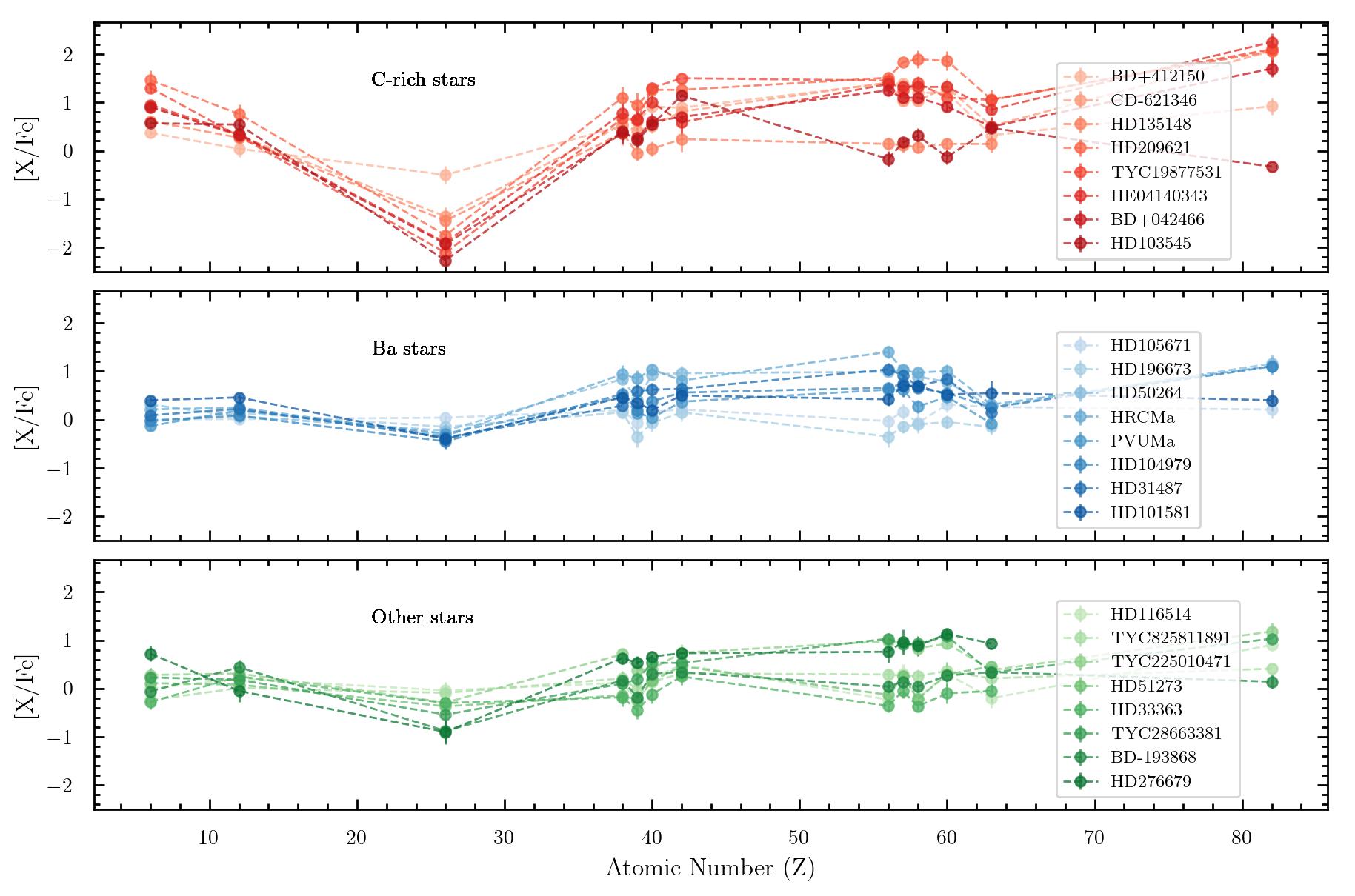}
    \caption{Abundance patterns for our sample of stars by stellar classification; carbon enhanced stars, Ba stars, and ``other'' stars.}
    \label{fig:abund_pattern}
\end{figure*}

Abundance patterns for our sample are displayed in Figure \ref{fig:abund_pattern}. Stars are categorized by class: carbon-enhanced stars in the top panel, barium stars in the middle panel,  and ``other'' stars in the bottom panel. We improve upon the patterns of all stars in our sample by adding Mo and, on average, we improve on existing abundance patterns by a factor of about 1.5 by adding new elements to the patterns. Stars in the bottom panel that are chemically interesting based on their carbon and heavy element enhancements are HD 116514, TYC 2250-1047-1, HD 51273, HD 276679, TYC 2866-338-1, and TYC 9244-8667-1. 

\paragraph{Carbon:} 
We derive the carbon abundance in our stars from synthetic spectrum analysis of the C$_2$ swan band at 5165 \AA$\;$and the CH band at 4313 \AA. The steep change in flux and the sensitivity of the C$_2$ band makes it a robust feature to precisely determine the carbon abundance in our high-resolution spectral data, and the CH band provides a verification check on the carbon abundance. By selection bias in our sample of metal-poor stars, we see enhancements in carbon at lower metallicities. For eight stars in our sample (CD-62 1346, HD 135148, HD 209621, TYC 1987-753-1, HE 0141-0343, BD+04 2466, HD 276679, and HD 103545) we observe large enhancements in [C/Fe] $\gtrsim +0.5$ dex. Many other features in the spectra are blended with molecular carbon lines, and understanding the carbon content ensures our atomic abundances are robust.

\paragraph{Magnesium:}
Magnesium abundances are computed by measuring the equivalent widths of the Mg lines at 5528 and 5711 \AA. These lines are weaker than the Mg b lines, and remain unsaturated. The 5711 \AA\; line is weak, and in cases where [Mg/Fe] $\sim$ 0.0, is indistinguishable from the continuum, including at higher metallicities. For our abundance sub-sample of stars, we find [Mg/Fe] values close to the solar value at higher metallicity ([Fe/H] $>$ -1) with some expected scatter. At lower metallicity, our results are in agreement with \citet{2019A&A...624A..19B}, following a flat $\alpha$-enhancement trend with some scatter.  

\begin{figure}
    \centering
    \includegraphics{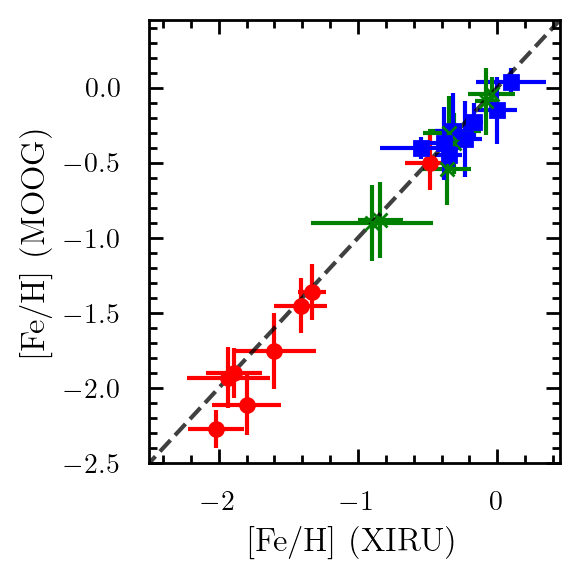}
    \caption{Comparison of [Fe/H] abundances between ionization and excitation balance from Xiru and spectral synthesis abundances using MOOG. There is good agreement between Xiru and MOOG, with only two metal poor stars slightly outlying from the 1-1 trend.}
    \label{fig:fe_comp}
\end{figure}

\paragraph{Iron:}
We determine spectroscopic iron abundances using the equivalent widths of Fe I and II lines determined by ARES and ionization and excitation balance in Xiru. We also compute Fe abundances from spectral synthesis in the region between 5180 and 5250 \AA, where many Fe I and II features exist. We compare the two values from Xiru and MOOG in Figure \ref{fig:fe_comp}. Xiru metallicity estimates in our abundance sample are in close agreement to the determined spectroscopic value. Lines of Fe I and Fe II should display the same abundance in the stellar spectrum; if the temperature or surface gravity is not well constrained, there may be a disagreement between the abundances in the two ionization states of iron. This effect is minimized when using ionization balance, for example in Xiru. While discrepancies in Fe I and II could be due to NLTE effects in some of the lines, particularly at lower metallicities, this provides another check on the Xiru $\mathrm{T_{eff}}$ and $\log g$ estimates.

\paragraph{s-Process:}
The abundances of Sr, Y, Zr, La, and Eu have been previously derived for a large part of our sub-sample, and we find our results in general agreement with the literature. There is visible scatter in the s-process element abundances in our sample, indicating different enrichment pathways in some of our targets. Some stars show large enhancements in s-process elements and are likely to have been enriched by an AGB companion, where some have lower enhancements and are likely to have been enriched by the parent molecular cloud. 

We observe a tight grouping in the [Mo/Fe] abundance in our abundance sub-sample in Figure \ref{fig:abund_grid}, with [Mo/Fe] generally between 0 and +1.0 dex, save for some upper limits. In Figure \ref{fig:abund_grid}, abundances of the heavy s-process elements Ba, La, Ce, and Nd exhibit more scatter in the metal-poor regime [Fe/H] $\lesssim -1.0$ \citep{2007A&A...476..935F, 2012arXiv1212.4147H, 2014A&A...568A..47H}.

The [Pb/Fe] space is more sparsely filled compared to other elements, as not all of our stars had Pb detections. The Pb line at 4057 \AA\;is blended with carbon and magnesium molecular features and can easily be washed out at lower S/N. For some of our metal-poor stars and Ba stars, we observe large enhancements in Pb, a key signature of the strong s-process in AGB stars. We observe one star HD 103545 with a low Pb abundance, and note that this star is a CEMP-no star, and the Pb abundance is only a limit.

Table \ref{tab:s_ls_hs} displays the metallicity, carbon enrichment, and relative s-process enhancement for our abundance sample, organised by nucleosynthetic classification (Ba, C-rich, unknown), and ordered by decreasing metallicity. Contributions from light ([ls], $\langle \mathrm{Sr, Y, Zr, Mo} \rangle$) and heavy ([hs], $\langle \mathrm{Ba, La, Ce, Nd} \rangle$) s-process enrichment, and the ratio of heavy-to-light s-process ([hs/ls]) elements are shown. Stars with positive [hs/ls] are likely to have been received this material from an AGB companion. 

\citet{2016MNRAS.459.4299D} provides a definition of a barium star as [s/Fe] $> +0.25$ and [Fe/H] $> -1$, based on high-resolution spectra. For simplicity we adopt this same definition; we confirm the large s-process enhancements and Ba-star nature of TYC 2250-1074-1, and we add the stars TYC 8258-1189-1 and TYC 2866-338-1 to this category. We suggest HD 276679 is a CH type star with large enhancements in carbon ([C/Fe] = 0.72) and s-process elements ([s/Fe] = 0.79).

\paragraph{Europium:} Computed abundances of the canonically r-process-produced element europium in Figure \ref{fig:abund_grid} show a general trend with little scatter, suggesting a more singular enrichment channel for r-process material in our stars that are predominantly s-process enriched. Since this is the case, we assert that our Mo enhancements are mainly from the s-process and not the r-process. Lower metallicity stars show [Eu/Fe] abundances typical of r-I type stars.

\begin{table}[t]
    \centering
    \caption{Ratios of s-process elements for each star in our abundance sub-sample, organized by increasing metallicity. }
    \tiny
    \begin{tabular}{l c c c c c c}
        \hline \hline
        Star 	 	& 	[Fe/H] 	 &  	[C/Fe] 	 & 	[s/Fe] 	 & 	 [ls] 	 & 	 [hs] 	 & 	[hs/ls] \\
        \hline \hline
        HD105671        & 	 0.10 	 & 	 0.01 	 & 	 0.12 	 & 	 0.13 	 & 	 0.09 	 & 	-0.03 \\
        HD196673        & 	 0.00 	 & 	-0.05 	 & 	-0.10 	 & 	-0.03 	 & 	-0.16 	 & 	-0.13 \\
        HD50264         & 	-0.17 	 & 	 0.31 	 & 	 0.89 	 & 	 0.79 	 & 	 0.92 	 & 	 0.14 \\
        HRCMa           & 	-0.23 	 & 	 0.21 	 & 	 1.02 	 & 	 0.91 	 & 	 1.10 	 & 	 0.19 \\
        PVUMa           & 	-0.32 	 & 	-0.13 	 & 	 0.37 	 & 	 0.25 	 & 	 0.49 	 & 	 0.24 \\
        HD104979        & 	-0.35 	 & 	-0.02 	 & 	 0.60 	 & 	 0.42 	 & 	 0.65 	 & 	 0.23 \\
        HD31487         & 	-0.38 	 & 	 0.09 	 & 	 0.70 	 & 	 0.54 	 & 	 0.87 	 & 	 0.34 \\
        HD101581        & 	-0.55 	 & 	 0.40 	 & 	 0.47 	 & 	 0.38 	 & 	 0.59 	 & 	 0.21 \\
        BD+412150       & 	-0.48 	 & 	 0.37 	 & 	 1.00 	 & 	 0.73 	 & 	 1.29 	 & 	 0.56 \\
        CD-621346       & 	-1.35 	 & 	 0.91 	 & 	 1.01 	 & 	 0.57 	 & 	 1.20 	 & 	 0.63 \\
        HD135148        & 	-1.41 	 & 	 0.61 	 & 	 0.13 	 & 	 0.15 	 & 	 0.12 	 & 	-0.03 \\
        HD209621        & 	-1.60 	 & 	 1.46 	 & 	 1.55 	 & 	 1.19 	 & 	 1.77 	 & 	 0.58 \\
        TYC19877531     & 	-1.80 	 & 	 1.30 	 & 	 1.27 	 & 	 1.03 	 & 	 1.30 	 & 	 0.28 \\
        HE04140343      & 	-1.89 	 & 	 0.95 	 & 	 1.18 	 & 	 0.76 	 & 	 1.34 	 & 	 0.58 \\
        BD+042466       & 	-1.93 	 & 	 0.90 	 & 	 0.89 	 & 	 0.48 	 & 	 1.09 	 & 	 0.60 \\
        HD103545        & 	-2.02 	 & 	 0.57 	 & 	 0.24 	 & 	 0.57 	 & 	 0.04 	 & 	-0.53 \\
        HD116514        & 	-0.04 	 & 	 0.00 	 & 	 0.18 	 & 	 0.14 	 & 	 0.04 	 & 	-0.10 \\
        TYC825811891    & 	-0.08 	 & 	-0.24 	 & 	 0.25 	 & 	 0.16 	 & 	 0.29 	 & 	 0.13 \\
        TYC225010471    & 	-0.31 	 & 	 0.28 	 & 	 0.79 	 & 	 0.56 	 & 	 0.93 	 & 	 0.37 \\
        HD51273         & 	-0.31 	 & 	 0.11 	 & 	 0.06 	 & 	 0.06 	 & 	 0.05 	 & 	-0.01 \\
        HD33363         & 	-0.35 	 & 	-0.38 	 & 	-0.17 	 & 	-0.13 	 & 	-0.22 	 & 	-0.09 \\
        TYC28663381     & 	-0.36 	 & 	-0.02 	 & 	 0.70 	 & 	 0.34 	 & 	 0.99 	 & 	 0.65 \\
        BD-193868       & 	-0.67 	 & 	-0.06 	 & 	 0.14 	 & 	 0.16 	 & 	 0.12 	 & 	-0.04 \\
        HD276679        & 	-0.90 	 & 	 0.72 	 & 	 0.79 	 & 	 0.64 	 & 	 0.93 	 & 	 0.30 \\
        \hline \hline
    \end{tabular}
    \tablefoot{Total s-process [s/Fe], light s-process (`ls' $\langle \mathrm{Sr, Y, Zr, Mo} \rangle$) and heavy s-process (`hs'  $\langle \mathrm{Ba, La, Ce, Nd} \rangle$) enrichment are compared.}
    \label{tab:s_ls_hs}
\end{table}

Uncertainties in our abundances vary slightly between spectral features and, on average, the atmospheric parameters contribute an uncertainty of $0.15$. With the average statistical uncertainty in our lines $0.10$, we arrive at a combined average uncertainty of $\langle \sigma_{tot} \rangle = 0.25$. Stellar atmospheric parameters can have a significant effect on computed abundances from stellar spectra. We perform a sensitivity study to determine the relative change in the abundance for a small change in each of the atmospheric parameters. For example $\partial\log\epsilon / \partial\xi$: by varying the microturbulence $\xi$ by $\pm 0.2$ km/s can alter the abundances of Sr, Ba, and La by $0.05$ dex, and in specific cases (e.g. strong lines or low surface temperatures) up to a maximum of $0.1$ dex. We find this acceptable, as it is within the combined uncertainties in our abundance computations, and assume similar variations from the microturbulence in other elements.

If the abundances of different ionization states of iron are not compatible within the spectra (i.e. $> 0.3$ dex separation), this may be indicative of NLTE effects in some of the lines, or a poorly fit atmospheric model. Broadening of spectral lines due to stellar rotation makes it difficult to compute abundances of weak lines. Spectral features will be blended together, effectively washing out finer details, even at high resolution. At high resolutions, we find rotation becomes an issue with rotational velocities $v\sin i \gtrsim 15$ km/s. If the rotation velocity is not known, we visually inspect the spectrum for rotational broadening effects. To this end, we chose stars for our abundance sample that have low, if measured, rotation velocities. 

\subsection{Orbital parameters}
Combining our computed heliocentric RVs with available literature data, we optimize binary orbits to estimate orbital and physical parameters of the star systems. Measured RVs from our observations are electronically available on the Milne-Center GitHub page\footnote{\url{https://github.com/Milne-Centre/Barium-Star-Repository/}}. We add low-error data points to stellar systems with our high-resolution snapshot spectra, with a typical improvement on the order of a factor of 5-10 compared to previous RV uncertainties. 

\begin{figure}[]
    \centering
    \includegraphics[width=0.91\linewidth]{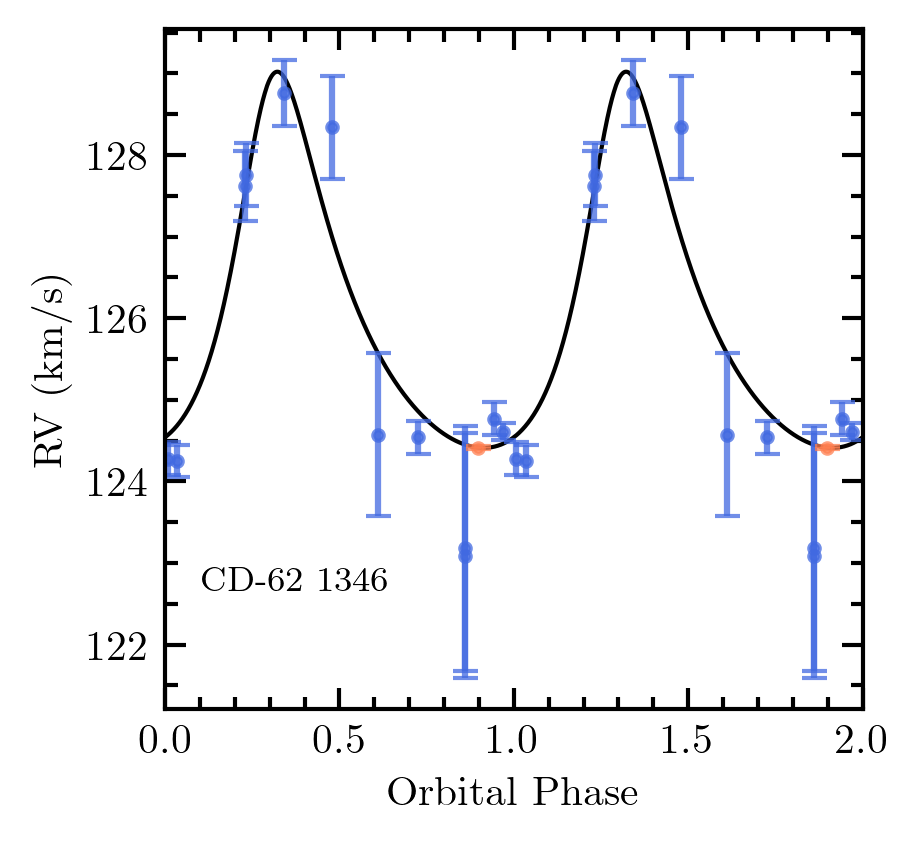} \\
    \includegraphics[width=0.89\linewidth]{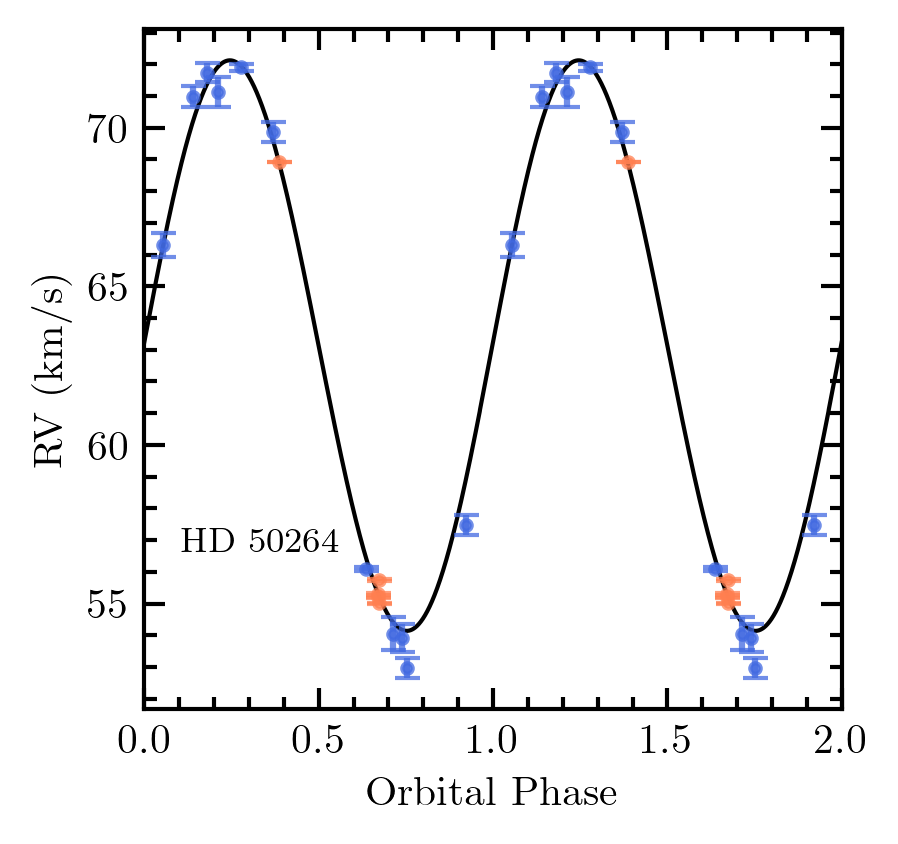} \\
    \includegraphics[width=0.90\linewidth]{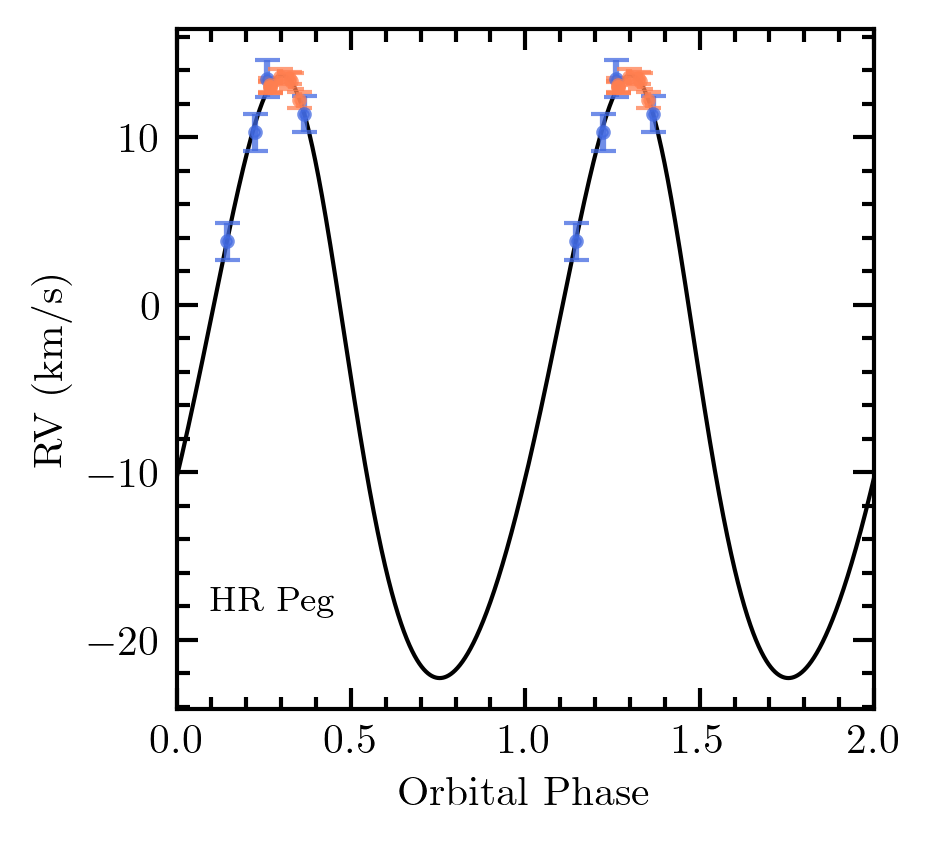}
    \caption{Phase-folded RV curves for CD-62 1346, HD 50264, and HR Peg. We use available literature RV data (blue) in conjunction with our own observations(orange). These three stars have few literature data points, and with our additions we have enough to characterize their orbits and constrain component masses.}
    \label{fig:RV_curves}
\end{figure}

Systems enriched by a previous AGB companion typically have long orbital periods and are old enough such that the AGB has faded to a white dwarf, providing time for orbital circularization. Most of our binary systems with abundance patterns indicative of AGB mass transfer have low eccentricity orbits and the orbital periods vary from a few hundred days to a few thousand days. Figure \ref{fig:RV_curves} displays phase-folded RV curves for selected systems CD-62 1346, HD 50264, and HR Peg. Notably, we estimate the orbit of CD-62 1346 for the first time, with our data in good agreement with the existing literature data points from CORrelation-RAdial-VELocities instrument (CORAVEL) and South African Large Telescope High Resolution Spectrograph (SALT-HRS).

The individual masses or the mass ratio are key outputs of our study, along with the standard binary orbital parameters. We note our masses are dependent on the assumption of the orbital inclination. Individual component masses can be approximated through the optimization, and are included in the table where available. A sub-sample of orbital and physical parameters from ELC are presented in Table \ref{tab:orbit_params_out}. The \# Obs. column includes the number of data points added to each system from our RV observations, and the \# Lit. column reads the number of literature data points collected and used in the orbital analysis. In HR CMa, the parameters with an asterisk are fixed values for the model fit. Here, $M_1$ refers to the visible component, and $M_2$ refers to the companion, typically a white dwarf. Our mass estimates are dependent on the inclination of the system, and we observe large scatter across the subset.

\section{Discussion}\label{sec:DISCUSSION}

\citet{2013ApJ...762...31P} investigated the impact of $^{12}$C fusion in massive stars, and the effect on the production of molybdenum. \citet{2014A&A...568A..47H} studied the nucleosynthetic origins of Mo in a sample of 52 stars, and sought correlations between Mo and other s- and r-process elements. Both groups found that Mo is highly convolved with other elements, and receives contributions from both the s- and r-processes. We find tighter correlations between Mo and light s-elements, and more scatter when comparing to heavy s-elements.

\begin{figure*}
    \centering
    \includegraphics[width=\linewidth]{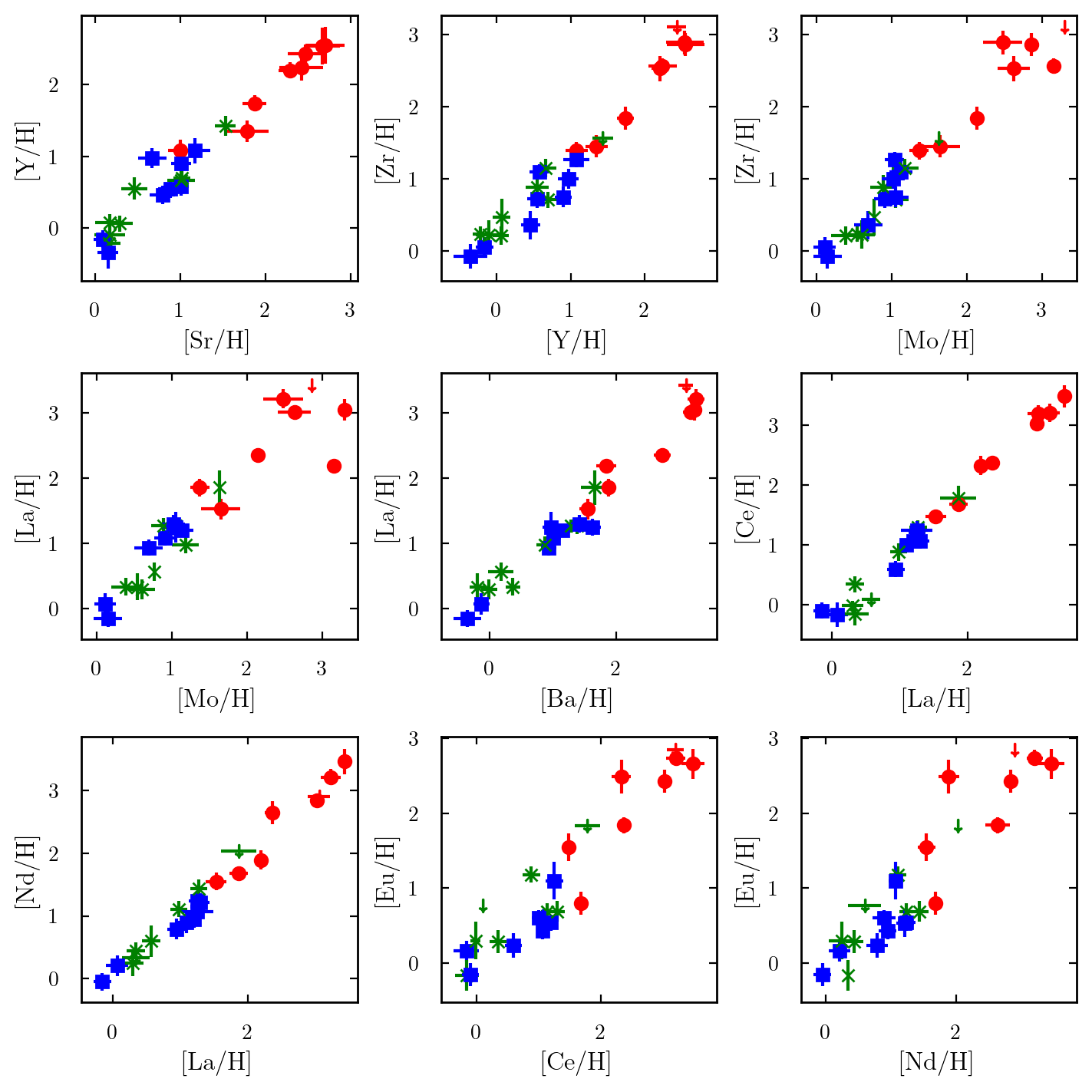}
    \caption{Elemental comparison between light- and heavy s-process elements. We compare [X/H] abundances to eliminate metallicity dependencies in the trends. We observe correlations in all shown comparisons.}
    \label{fig:elem_elem_comparison}
\end{figure*}

We compare our stellar abundance distributions to each other in [X/H] space to eliminate metallicity dependencies in Figure \ref{fig:elem_elem_comparison}, and look for correlations in abundance space. Very tight relations with slope of one indicate co-production of the elements; scatter in these relations or deviation from the slope-one line is indicative of multiple nucleosynthetic production pathways. We expect tight relations within the light and heavy s-process element groups (for example Y and Sr, or Ba and La), and between both light- and heavy s-elements. 

We observe general correlations in all panels of \ref{fig:elem_elem_comparison}; the elements we study are formed at least partially through the s-process. The average of our Mo distribution ($\sim0.57$) is between that of the heavy s-elements ($\sim0.66$) and the light s-elements ($\sim0.45$). There exist correlations between Mo and Zr, and Mo and La; one each of the light and heavy s-process elements, although there is scatter at higher Mo / Zr / La abundances, which also trend with decreasing metallicity. \cite{2014A&A...568A..47H} investigated the production of Mo in stars, and found multiple pathways to increased Mo abundances. At higher [Fe/H], Mo may be produced by the p-process, and at lower metallicities Mo correlates directly with Sr and Zr, pointing towards contributions from the early weak s-process. A corner plot of the canonical s-process elements is available in Figure \ref{fig:abundance_corner}.

We find trends with scatter when comparing the heavy s-elements Ce and Nd with the r-process element Eu in Figure \ref{fig:elem_elem_comparison}. As Ce is mainly produced by the s-process ($\sim80\%$ s vs $\sim20\%$ r); reduced scatter in the Eu vs Nd panel gives weight to the multiple formation pathways of Nd ($\sim50\%$ s vs $\sim50\%$ r), with Eu being produced $\sim94\%$ by the r-process. 

We visually compare our abundances [X/Fe] to the galactic nucleosynthesis analysis of \citet{2020ApJ...900..179K}, and find similar trends. Our Mg abundances follow the same general $\alpha$-element trend with near-solar values at higher metallicities and slight enhancements at lower metallicities. The heavy element trends also follow, with increased scatter at lower metallicities. 

\subsection{FRUITY models}
We compare our abundance measurements to the FRUITY yields \citep{2011ApJS..197...17C} to investigate the origins of our observed abundance patterns. The FRUITY database contains around $120$ models that range in initial mass from 1.3 - 6 M$_{\odot}$, metallicities from Z = 0.00002 to 0.03 ([Fe/H] = -2.85 to 0.32), and initial rotational velocities of 0, 10, and 30 km/s. FRUITY allows for the free creation of the $^{13}$C pocket by parameterization of physical mixing processes through the thermal pulses. Since the FRUITY models are of AGB surface abundances, this material is diluted upon accretion onto a binary companion through convective and thermohaline mixing processes. We approximate the mixing of the stellar abundances and FRUITY abundances using a prescription identical to \cite{2023A&A...672A.143D}. The material accreted onto the stellar surface is diluted such that 

\begin{equation}\label{eqn:dilute}
    \left[ \frac{\mathrm{X}}{\mathrm{Fe}} \right] = \log_{10} \left[ (1-\delta) \times 10^{[X/Fe]_{ini}} + \delta \times 10^{[X/Fe]_{AGB}} \right],
\end{equation}

\noindent where [X/Fe]$_{ini}$ is the initial abundance of element X, and [X/Fe]$_{AGB}$ is the final surface abundance of the AGB model, and $\delta$ is the dilution factor. Higher dilution factors imply the observed stellar envelope is less mixed, and mostly composed of AGB material. A dilution factor of zero (0) results in a flat abundance profile, and no indication of heavy element enhancement from an AGB star.

\begin{figure*}[]
    \centering
    \includegraphics[width=\linewidth]{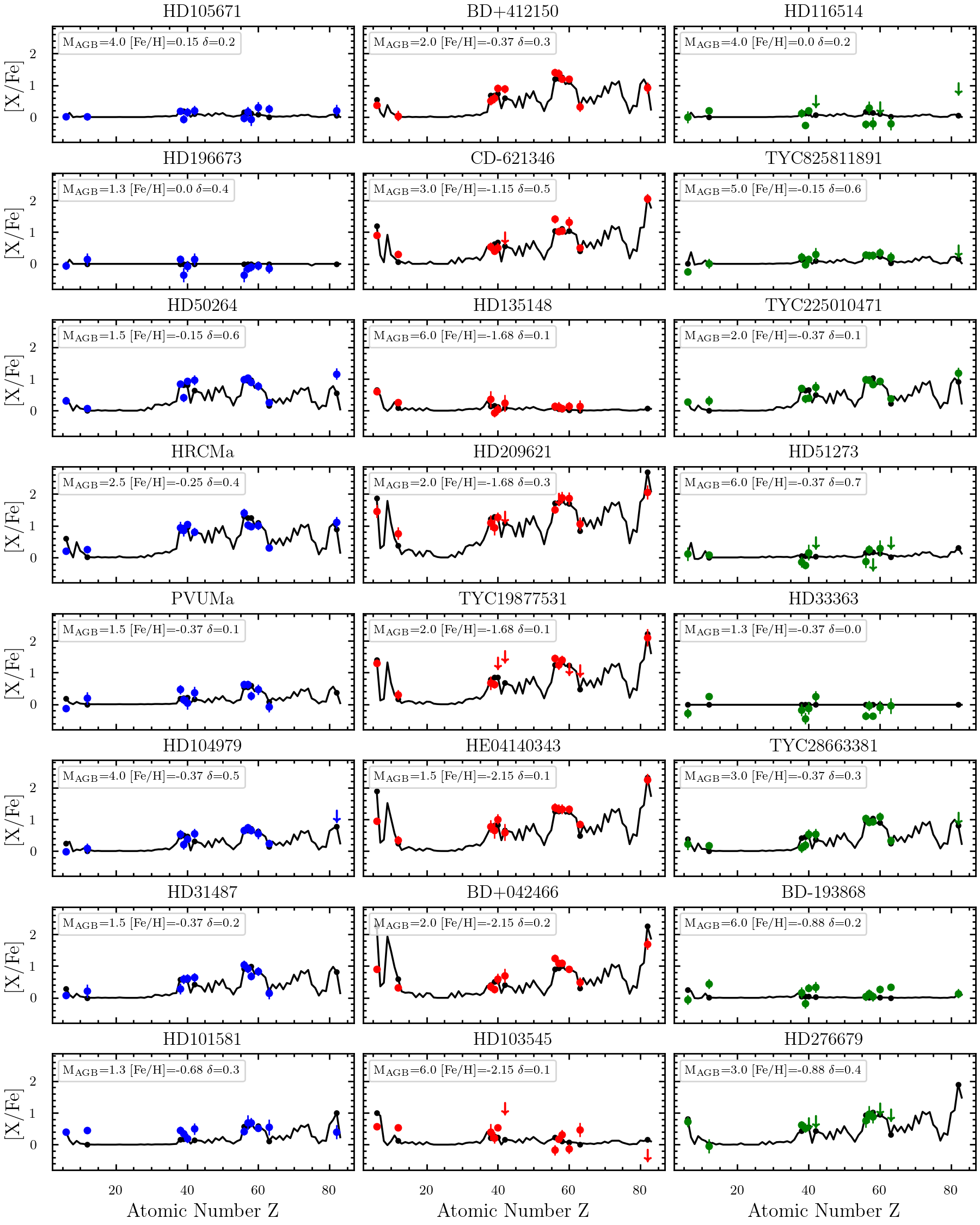}
    \caption{FRUITY models compared to computed abundances for our high-quality abundance sample, organized by decreasing metallicity. Blue data points correspond to Ba stars, red to C-enriched stars, and green to ``other'' stars. Inverted triangles in the plots are upper limits from our abundance computations. FRUITY model data is shown in black.}
    \label{fig:FRUITY_compare}
\end{figure*}

We find the model to best fit our observed stellar abundances using a least-squared fitting method by comparing the abundances measured in our sample to those produced by the model AGB stars within a range of masses, metallicities, initial rotation rates, $^{13}$C pocket formation, and dilution factors. For each star, the abundance pattern and best fit model are displayed in Figure \ref{fig:FRUITY_compare}, organized by chemical composition and decreasing metallicity, with the most metal rich stars in each group at the top of the plot. Colors in Figure \ref{fig:FRUITY_compare} correspond to to stellar classification, as in other figures in this work.

Most of the Ba, CH, and CEMP-s stars in our abundance sub-sample find good fits to the FRUITY AGB yield models with s-process signatures, visible in the two (or three) peaks around Sr and Ba (and Pb where available). The weak Ba stars HD 105671 and HD 196673 show less pronounced peaks around Sr and Ba. The CEMP-no stars HD 135148 and HD 103545 show flat abundance patterns with an AGB mass of  $M_{AGB} = 6.0$, indicating, that these stars have not been enriched by the s-process from a reasonable AGB companion. Some of our stars are metal-poor, where chemical enrichment processes at low metallicities may operate differently than those that produced the solar abundance pattern. The nucleosynthetic i-process is an alternative option to explain discrepancies in these patterns.

The amount of material transferred from an AGB star depends on the orbital separation, where closer binaries will likely experience a higher mass-transfer efficiency. However, if the orbital separation is too small, mass transfer will occur through Roche-Lobe overflow (RLOF), which may result in a different chemical enrichment signature; this scenario more complex and out of the scope of this study. However, one star in this work (HD 116514) may be the result of RLOF mass transfer - see Section \ref{sec:apx_other}. We identify 10 stars in our sample with longer orbital periods ($>$100 days) to model using the \texttt{STARS} stellar evolution code in a follow-up paper, and 2 stars with shorter orbital periods that may be RLOF systems. 

While we have focused this study on s-process signatures of low mass AGB stars, the s-process also occurs in rapidly-rotating massive stars \citep{2012A&A...538L...2F}. Rotation rates required to induce internal mixing and the weak s-process in these massive stars are on the order of half of the critical rotation velocity. These massive stars may have polluted the ISM in regions of the galaxy where some of our stars have formed, that show mild enhancements in [s/Fe] but do not fit well to FRUITY yields with the large double-peaked signature indicative of AGB mas transfer. 

\subsection{RV variability}
To characterize binary orbits of stars in our sample, we use the ELC program to model systems with sufficient RV data; generally six to eight data points if they are well spread across the orbit. Collected RV data is generally of good quality, within $\sim0.5$ km/s precision with ChETEC-INFRA instruments, and within $\sim0.05$ km/s with FIES and FEROS on average. 

To estimate average uncertainties in our radial velocity measurements from with TNA telescopes, we use Equation (1) in \citet[][]{2020PASP..132c5002K}: 
\begin{equation}
    \sigma_{RV} = C\;\times\;(S/N)^{-1}\;\times\;\Delta\lambda^{-0.5}\;\times\;R^{-1.5},
\end{equation}
\noindent where for the OES instrument, the wavelength range $\Delta\lambda = 5900$ \AA, the resolution $R \sim 50000$, and an instrument-specific constant $C = 6.5\times10^{12}$. For a S/N of $\sim 15$, deemed adequate for RV monitoring, an accuracy of $\approx 500\;m/s$ or $0.5\;km/s$ can be achieved. This expression is for a solar-like star, but we find it adequate for our investigation of metal-poor stars and giants alike. The standard deviations between spectral orders from the TNA instruments are close to these values, and we find the approximation acceptable. 

Our measured RVs are in good agreement with literature data, and systematic offsets have been corrected with observations RV standard stars and converting to HJD. Velocity variability for known binaries is within the expected ranges, and this gives us confidence in observed variability in binary candidates. Computed mass functions $f(m)$ in Table \ref{tab:orbit_params_out} are generally sensible, and in good agreement with literature values. 

The s-process enhanced stars TYC 8258-1189-1, TYC 2250-1047-1, TYC 2866-338-1, HD 276679, and the CEMP-s star TYC 1987-753-1 show promising abundance patterns (Figure \ref{fig:abund_pattern}) for s-process enrichment from an AGB companion, but do not show appreciable RV variability with only a few time-series data points. We have scheduled follow up observations to characterise the binarity of these targets. We do not detect appreciable radial velocity variation in the CEMP-no star HD 103545 or in the mild Ba star HD 101581. 

We find that stars with sufficient RV data in our abundance sub-sample generally have eccentricities $\lesssim0.15$, even for the longest period systems. This is expected for older systems, where they have had enough time to circularize their orbits. This is further evidence that Ba, CH, and CEMP-s stars obtain their heavy element signatures from an AGB companion where the system has had enough time to evolve the AGB to a white dwarf. 

\subsection{Stellar masses and ages}

\begin{figure}[]
    \centering
    \includegraphics[width=\linewidth]{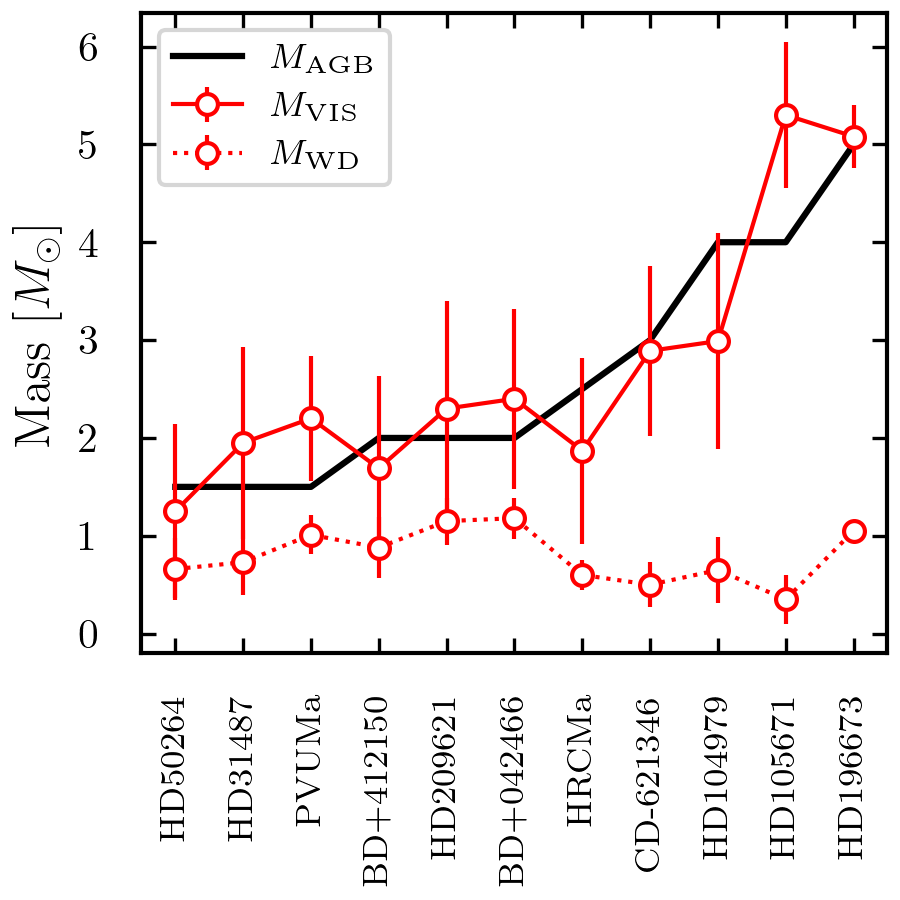}
    \caption{Estimated masses for AGB donor stars from FRUITY (black line), and dynamically derived visible (solid red line) and white dwarf (dashed red line) masses from ELC. }
    \label{fig:AGB_vis_WD_mass_comp}
\end{figure}

The masses of the AGBs that produced the s-process elements should be roughly greater than or equal to the initial mass of the observable star for evolutionary reasons, with some allowances for accreted mass. The AGB masses from FRUITY are compared to the mass estimates from the binary orbit modelling in Figure \ref{fig:AGB_vis_WD_mass_comp}. As the AGB evolves and loses mass, the now faint white dwarf should have a smaller mass than the visible star, and should not be more massive than the Chandrasekhar mass limit for white dwarfs, 1.44 M$_{\odot}$. In Figure \ref{fig:AGB_vis_WD_mass_comp}, we observe white dwarf masses (m$_{WD}$ or M$_2$, gold line) less than those of the initial AGB mass from FRUITY (dark red line), and also less than those of the visible components (m$_{Vis}$ or M$_1$, orange line). For the stars HD 31487, PV UMa, HD 209621, BD+04 2466, and HD 105671, the visible component is slightly more massive than the AGB fit from FRUITY, although this is within the error bars of the estimated visual mass. 

With our estimated stellar masses and atmospheric parameters, we can approximate the age of our program stars. Using PARSEC isochrones \citep{2012MNRAS.427..127B, 2014MNRAS.444.2525C, 2015MNRAS.452.1068C, 2014MNRAS.445.4287T, 2017ApJ...835...77M} for metallicities comparable to those in our abundance sub-sample, we investigate the ages for eight of our best-fit stellar systems: HD 50264, HR CMa, HD 104979, BD+41 2150, HD 31487, HD 209621, CD-62 1346, and BD+04 2466. We choose these systems because they display strong s-process enhancement, and show good agreement between AGB mass, dynamical mass, and white dwarf masses from FRUITY and ELC. We initialize our isochrones with an IMF from \citet{2013pss5.book..115K}, which corrects for unresolved binaries. We compare our stars to the isochrones based on our dynamical mass estimates from ELC, Xiru estimates of effective temperature and surface gravity, and extinction-corrected absolute G magnitudes (G$_0$) computed from \emph{Gaia} DR3 apparent G-magnitudes and parallaxes.

For a given metallicity, we determine on which isochrone our star sits in T$_{eff}$-mass space, $\log g$-mass space, and G$_0$-mass space, and compare these results to the standard HR diagram of T$_{eff}$-G$_0$ space. We find the best matching isochrone point by minimizing the distance between our data and the isochrones in a least-squares fitting routine. Approximated ages of our systems can be found in Table \ref{tab:parsec_ages}. We find the ages determined from our estimated masses to be in general agreement with those determined in the HR-diagram space. However, for more metal poor stars, we find a systematic offset of about $0.7$ dex towards younger ages when using our mass estimates; one would not expect a metal poor star such as BD+04 2466 to be less than $10^9$ years old. However, one would not expect a two-solar-mass star to be older than about 1 Gyr, and for metal-poor stars we find the ages determined from our dynamical mass estimates to be more robust.

\begin{table*}[]
    \centering
    \caption{Stellar metallicities, masses, and estimated ages.}
    \begin{tabular}{c c c c c c c c c}
    \hline \hline
        Star & [Fe/H] & $\sigma_{\mathrm[Fe/H]}$ & Mass ($M_{\odot}$) & $\sigma_{M}$ & log(Age$_{M}$) & $\sigma$(Age$_{M}$)& log(Age$_{HR}$) & $\sigma$(Age$_{HR}$) \\
    \hline
        HD 50264	& -0.17 & 0.07 & 1.25	& 0.85  & 9.55	& 0.11	& 9.71	& 0.33 \\
        HR CMa	    & -0.23 & 0.12 & 1.87	& 0.95  & 9.12	& 0.07	& 9.47	& 0.42 \\
        HD 104979	& -0.35 & 0.10 & 2.99	& 1.10  & 9.36	& 0.16	& 9.50  & 0.33 \\
        HD 31487	& -0.38 & 0.14 & 1.95	& 0.98  & 9.19	& 0.07	& 9.42	& 0.47 \\
        BD+41 2150	& -0.48 & 0.18 & 1.69	& 0.94  & 9.19	& 0.08	& 9.42	& 0.47 \\
        CD-62 1346	& -1.33 & 0.10 & 2.89	& 0.87  & 8.52	& 0.14	& 9.79	& 0.30 \\ 
        HD 209621	& -1.60 & 0.40 & 2.30	& 1.10  & 8.63	& 0.49	& 9.45	& 0.36 \\
        BD+04 2466	& -1.93 & 0.30 & 2.40	& 0.92  & 8.49	& 0.51	& 9.55	& 0.33 \\
    \hline \hline
    \end{tabular}
    \tablefoot{Ages are based on fitting stellar parameters on PARSEC isochrones (HR) and using the dynamical stellar mass estimates from ELC (M).}
    \label{tab:parsec_ages}
\end{table*}

\begin{figure*}
    \centering
    \includegraphics[width=\linewidth]{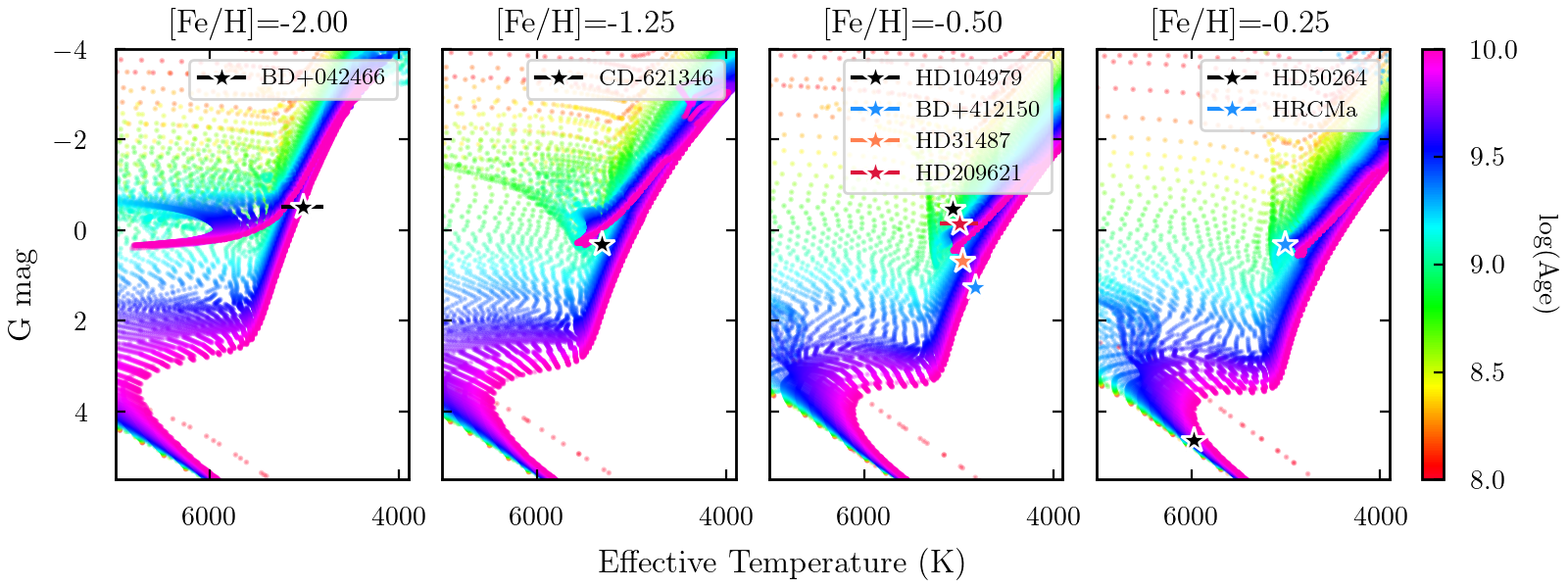}\\
    \caption{Isochrones for the best fit stars for which we have estimated dynamical masses in compatible mass and metallicity bins.}
    \label{fig:stellar_ages}
\end{figure*}

For these eight selected stars, stellar ages from the HR diagram are over $10^9$ years, with the oldest around $6.2\times10^9$ years. The ages derived from mass estimates are between $3 .3\times10^8$ and $3.5\times10^9$ years. In Figure \ref{fig:stellar_ages} we plot our stars against isochrones in the HR-Diagram parameter space (magnitude vs temperature), organized by metallicity. The color bar represents the age of the isochrones, scattered in the background ranging from $10^8$ to $10^{10}$ years, with our stars plotted on top.

The star BD+04 2466 sits on the red giant branch, but the low surface gravity of $\log g = 1.68$ may suggest the possibility of being an AGB star; uncertainties in the temperature overlap with some of the asymptotic giant branch. CD-62 1346 lies between the RGB and the red clump. The low surface gravity may indicate advanced evolution from the RGB towards the horizontal branch (HB). HD 104979 is likely on the RGB, or advancing onto the HB in the HR diagram. The surface gravity $\log g = 2.08$ does not suggest this star is an AGB star. BD+41 2150 is definitively ascending the RGB, undergoing hydrogen shell fusion. While we find a surface gravity of $<$ 2, the error bars are not small and we do not suggest this star is an AGB. HD 31487 is also on the RGB, in a similar evolutionary state as BD+41 2150. HD 209621 sits in a similar place to HD 104979, somewhere between the RGB and HB phases. The lower surface gravity of $\log g = 1.64$ indicates advanced evolution, and could soon initiate He burning. HR CMa sits in the red clump region of the HR diagram, possibly undergoing helium fusion in the core before ascending the AGB - the lower surface gravity of $\log g = 1.98$ indicates advanced RGB or HB evolution. HD 50264 is a dwarf star with a high surface gravity $\log g = 4.24$, and lies on the main sequence at the bottom of the HR diagram. 

\begin{figure}
    \centering
    \includegraphics[width=\linewidth]{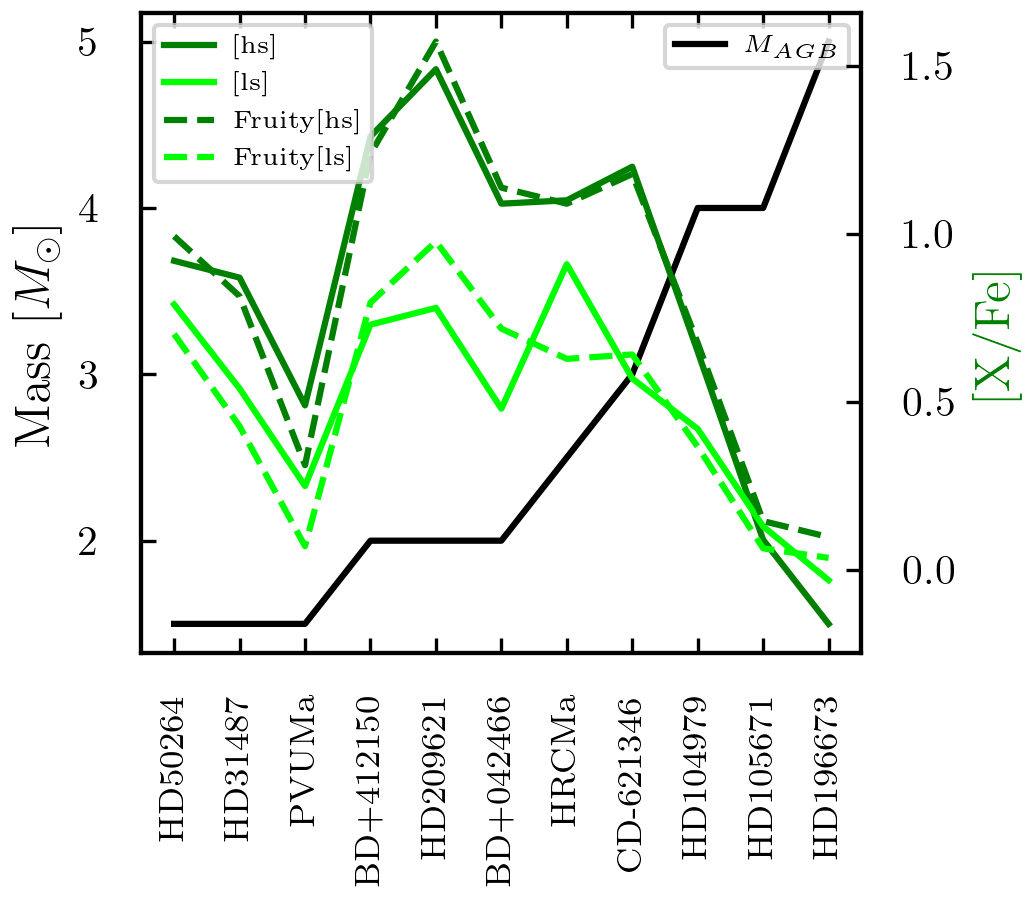}
    \caption{AGB donor mass (dark red) compared with observed s-process enrichment (shades of green).}
    \label{fig:AGB_mass_vs_s_proc}
\end{figure}

In Figure \ref{fig:AGB_mass_vs_s_proc} we compare the estimated AGB donor mass from FRUITY (red line) and the level of s-process enrichment from our abundance analysis (green lines). With the exception of the mild Ba star PV UMa, we observe generally more s-process material from low- and intermediate-mass AGB stars, with masses between $2-3$ M$_{\odot}$. The exception of PV UMa could be the mild Ba nature of the star, but with a much closer orbital separation and a much shorter orbital period, this may signal that a combination of RLOF and common envelope evolution results in an overall reduced accretion mass and therefore less material transferred, compared to the wider binaries which are more likely to have transferred overall more material via strong AGB winds. In the extreme cases of m$_{AGB} \sim 5 - 6$ M$_{\odot}$ such as HD 105671 and HD 196673, we observe little s-process enrichment. 

\section{Conclusions}\label{sec:CONCLUSION}

We observe spectroscopic binaries with high-resolution spectrographs to trace AGB nucleosynthesis patterns in the now-visible extrinsic binary companions displaying s-process element enhancements. Using atmospheric parameters estimated with Xiru and interpolated ATLAS9 / Kurucz atmospheric models, we compute 1D-LTE photospheric abundances in a sample of $24$ stars with MOOG. Adding Mo to the abundance pattern is useful in determining the initial AGB mass in binary systems polluted with s-process material. We see similar trends between both light and heavy s-process elements and Mo, and note correlations in elemental abundances produced by the s-process.

Comparing our computed abundances to FRUITY model AGB yields, we investigate systems that are likely to have been polluted by AGB material, and estimate AGB initial masses. We optimize binary orbits to further investigate the estimated stellar masses of our sample with constraints on the mass from the FRUITY models and the literature. We find general agreement in the observed stellar masses, initial AGB masses, and inferred white dwarf masses. For stars with good abundance fits to the FRUITY models, we find a range of initial AGB masses corresponding to enrichment in s-process material, and observe a general trend between s-process enhancement and donor AGB mass, where low- and intermediate-mass AGB stars produce more [hs] material compared to higher mass AGB stars.

We make note that BD+04 2466 could be approaching the AGB phase with low surface gravity and advanced position on the HR diagram. PV UMa is a weak Ba star with a short period of 80 days, and HD 116514 with an even shorter period of 5 days. The short period systems HD 116514 and PV UMa could be the results of RLOF accretion and common envelope evolution. We confirm the classification of TYC 2250-1047-1 as a Ba star based on its chemical composition, and suggest adding the star HD 276679 to the Ba star category, and TYC 8258-1189-1 to the mild Ba star category based on their chemical composition. Our evolutionary analysis reveals that HR CMa is a red-clump star, potentially already undergoing helium burning.

Using the ChETEC-INFRA TNA telescopes and the MPG instruments FIES and FEROS, we continue to observe RVs of our sample and compute abundances for our chemically peculiar stars. As spectra of the same targets are collected during followup RV observations, the spectra will be co-added for improved signal-to-noise ratios, and abundance analysis can be done on the full sample. 

\begin{acknowledgements}
      This project has received funding from the European Union’s Horizon 2020 research and innovation program under grant agreement No 101008324 (ChETEC-INFRA). We graciously thank ChETEC-INFRA TNA program for providing the framework and infrastructure for the access to the telescopes and instruments used to accomplish data acquisition. 
      Results based on observations made using the 1.65m Ritchey-Chretien telescope at Moletai Astronomical Observatory. These results are based on observations collected by 2m Perek Telescope of Ond\v{r}ejov observatory, run by Astronomical Institute of the Czech Academy of Sciences. The Astronomical Institute Ondřejov is supported by the project RVO:67985815. These results are based observations with the 2m Rozhen Ritchie-Chretien-Coud\'{e} telescope at the National Astronomical Observatory – Rozhen, operated by the Institute of Astronomy, Bulgarian Academy of Sciences. Based on observations made with the Nordic Optical Telescope, owned in collaboration by the University of Turku and Aarhus University, and operated jointly by Aarhus University, the University of Turku and the University of Oslo, representing Denmark, Finland and Norway, the University of Iceland and Stockholm University at the Observatorio del Roque de los Muchachos, La Palma, Spain, of the Instituto de Astrofisica de Canarias. 
      Further support in funding comes from the State of Hesse within the Research Cluster ELEMENTS (Project ID 541 500/10.006). 
      We acknowledge the support of the Data Science Group at the Max Planck Institute for Astronomy (MPIA) and especially Iva Momcheva, Morgan Fouesneau, and James Davies for their invaluable assistance in analyzing the data and developing the software for this research paper.
\end{acknowledgements}

%
%

\bibliographystyle{aa}
\bibliography{references}{}

\begin{appendix} 

\section{Stellar atmospheric parameters}

\begin{table*}[]
    \centering
    \caption{Estimated atmospheric parameters for our abundance sample using Xiru, compared to literature values.}
    \begin{tabular}{l c c c c c c c c c}
        \hline \hline
        Star & $T_{eff}$ & $\sigma_T$ & $\log(g)$ & $\sigma_{\log{g}}$ & [Fe/H] & $\sigma_{\mathrm{[Fe/H]}}$ & $\xi$ & $\sigma_{\xi}$  & Ref.\\
        \hline \hline
		HD 105671       & 4700 & 200 & 4.50 & 0.25 &  0.10 & 0.25 & 1.19 & 0.28	& \\
                        & 4617 &     & 4.55 &      &  0.07 &      &      &      & (1) \\
		HD 196673       & 4825 &  66 & 0.94 & 0.37 &  0.00 & 0.14 & 1.81 & 0.13	& \\
                        & 4914 &     & 2.50 &      &  0.12 &      &      &      & (2) \\
		HD 50264        & 5972 &  45 & 4.24 & 0.12 & -0.17 & 0.07 & 0.86 & 0.08	& \\
                        & 5900 &     & 4.60 &      & -0.13  &      &      &     & (3) \\
		HRCMa           & 5063 &  75 & 1.98 & 0.36 & -0.23 & 0.12 & 1.82 & 0.14	& \\
                        & 4822 &     & 2.40 &      & -0.23 &      &      &      & (4) \\
		PV UMa          & 5058 &  74 & 2.28 & 0.34 & -0.32 & 0.13 & 2.00 & 0.13	& \\
                        & 5050 &     & 2.50 &      & -0.13 &      &      &      & (2) \\
		HD 104979       & 5007 &  53 & 2.08 & 0.20 & -0.35 & 0.10 & 1.65 & 0.09	& \\
                        & 4933 &     & 2.68 &      & -0.33 &      &      &      & (1) \\
		HD 31487        & 4964 &  80 & 1.94 & 0.40 & -0.38 & 0.24 & 2.08 & 0.13	& \\
                        & 4960 &     & 3.11 &      & -0.04 &      &      &      & (5) \\
		HD 101581       & 5366 &  51 & 3.82 & 0.64 & -0.55 & 0.29 & 3.18 & 0.12	& \\
                        & 4738 &     & 4.46 &      & -0.52 &      &      &      & (1) \\
		BD+41 2150      & 4828 &  84 & 1.73 & 0.51 & -0.48 & 0.18 & 1.77 & 0.16	& \\
                        & 4707 &     & 2.16 &      & -0.67 &      &      &      & (6) \\
		CD-621346       & 5318 &  78 & 1.40 & 0.28 & -1.35 & 0.10 & 1.81 & 0.10 & \\
                        & 5300 &     & 1.70 &      & -1.57 &      &      &      & (7) \\
        HD 135148       & 4573 & 130 & 1.26 & 0.60 & -1.41 & 0.24 & 1.95 & 0.21	& \\
                        & 4237 &     & 0.66 &      & -1.89 &      &      &      & (1) \\        
		HD 209621       & 4850 & 443 & 1.64 & 0.33 & -1.60 & 0.40 & 2.06 & 0.45 & \\
                        & 4740 &     & 1.75 &      & -2.00 &      &      &      & (8) \\
		TYC 1987-753-1  & 6047 & 200 & 4.00 & 0.25 & -1.80 & 0.25 & 0.98 & 0.20	& \\
                        & 6126 &     & 3.45 &      & -2.09 &      &      &      & (9) \\
		HE 0414-0343    & 5204 & 118 & 2.00 & 0.62 & -1.89 & 0.15 & 0.80 & 0.14 & \\
                        & 4863 &     & 1.25 &      & -2.24 &      &      &      & (1) \\
		BD+04 2466      & 5021 & 220 & 1.68 & 0.15 & -1.93 & 0.30 & 1.66 & 0.20	& \\
                        & 4991 &     & 1.43 &      & -1.97 &      &      &      & (1) \\
		HD 103545       & 4960 & 121 & 1.51 & 0.79 & -2.02 & 0.17 & 2.14 & 0.19 & \\
                        & 4807 &     & 1.70 &      & -1.99 &      &      &      & (1) \\
        HD 116514       & 6247 & 127 & 4.25 & 0.58 & -0.04 & 0.17 & 2.93 & 0.18	& \\
                        & 5687 &     & 4.11 &      &  0.02 &      &      &      & (10) \\        
		TYC 8258-1189-1 & 5158 &  63 & 2.64 & 0.23 & -0.08 & 0.08 & 1.20 & 0.12 & \\
                        & 5171 &     & 2.77 &      & -0.14 &      &      &      & (11) \\
        TYC 2250-1047-1 & 5853 &  97 & 3.55 & 0.56 & -0.31 & 0.19 & 2.13 & 0.17	& \\
                        & 5335 &     & 3.71 &      & -0.55 &      &      &      & (12) \\        
		HD 51273        & 6412 & 103 & 3.46 & 0.43 & -0.31 & 0.12 & 1.75 & 0.17	& \\
                        & 6249 &     & 3.87 &      & -0.79 &      &      &      & (13) \\
        HD 33363        & 4947 &  61 & 2.07 & 0.42 & -0.35 & 0.18 & 2.25 & 0.12	& \\
                        & 4660 &     & 2.92 &      & -0.06 &      &      &      & (14) \\        
		TYC 2866-338-1  & 4837 &  78 & 2.78 & 0.35 & -0.51 & 0.13 & 1.86 & 0.12	& \\
                        & 4475 &     & 1.91 &      & -0.60 &      &      &      & (6) \\
		BD-193868       & 4402 &  95 & 1.55 & 0.45 & -0.67 & 0.13 & 1.32 & 0.18 & \\
                        & 4451 &     & 1.39 &      & -0.54 &      &      &      & (15) \\
		HD 276679       & 5934 & 115 & 3.18 & 0.50 & -0.90 & 0.18 & 1.05 & 0.18	& \\
                        & 6467 &     & 4.12 &      & -1.31 &      &      &      & (13) \\
        \hline \hline
    \end{tabular}
    \tablebib{(1) \cite{2022A&A...663A...4S} (2) \cite{2019A&A...626A.127J} (3) \cite{2019MNRAS.486.3266P}
    (4) \cite{2019ApJ...879...69T} (5) \cite{2018A&A...618A..32K} (6) \cite{2023MNRAS.524.1855Z}
    (7) \cite{2012A&A...543A..58P} (8) \cite{2021A&A...645A..61K} (9) \cite{2021ApJ...907...10L}
    (10) \cite{2020AJ....160..120J} (11) \cite{2018MNRAS.478.4513B} (12) \cite{2021A&A...654A.140K}
    (13) \cite{2023A&A...674A...1G} (14) \cite{2009A&A...504..829G} (15) \cite{2020AJ....160...83S}}
    \label{tab:atmos_params}
\end{table*}

\FloatBarrier

\section{Stellar abundances}

\begin{sidewaystable*}
    \caption{Computed abundances [X/Fe] for our sub-sample of stars, compared to literature values.}
    \scriptsize
    \centering
    \begin{tabular}{llllllllllllll}
        \hline \hline
        Star & [Fe/H] & C & Mg & Sr & Y & Zr & Mo & Ba & La & Ce & Nd & Eu & Pb \\
        \hline \hline
        HD & 0.10 $\pm$ 0.25 & 0.01 $\pm$ 0.09 & 0.01 $\pm$ 0.11 & 0.19 $\pm$ 0.11 & -0.06 $\pm$ 0.12 & 0.16 $\pm$ 0.14 & 0.21 $\pm$ 0.14 & -0.03 $\pm$ 0.13 & 0.17 $\pm$ 0.16 & -0.07 $\pm$ 0.2 & 0.31 $\pm$ 0.16 & 0.26 $\pm$ 0.14 & 0.21 $\pm$ 0.18 \\
        105671    & -0.03$^8$ & - - & 0.03$^8$ & 0.03$^8$ & 0.01$^8$ & 0.06$^8$ & - - & -0.10$^8$ & - - & 0.00$^8$ & - - & - - & - - \\
        \hline  
        HD & 0.00 $\pm$ 0.14 & -0.05 $\pm$ 0.13 & 0.14 $\pm$ 0.2 & 0.15 $\pm$ 0.12 & -0.35 $\pm$ 0.23 & -0.07 $\pm$ 0.17 & 0.15 $\pm$ 0.19 & -0.35 $\pm$ 0.22 & -0.15 $\pm$ 0.13 & -0.1 $\pm$ 0.14 & -0.05 $\pm$ 0.14 & -0.15 $\pm$ 0.16 & - - \\
        196673    & 0.12$^{2}$ & - - & - - & - - & 0.00$^{2}$ & 0.25$^{2}$ & - - & - - & 0.39$^{2}$ & 0.31$^{2}$ & - - & - - & - - \\
        \hline  
        HD & -0.17 $\pm$ 0.07 & 0.31 $\pm$ 0.09 & 0.07 $\pm$ 0.11 & 0.84 $\pm$ 0.09 & 0.41 $\pm$ 0.13 & 0.93 $\pm$ 0.10 & 0.96 $\pm$ 0.15 & 0.99 $\pm$ 0.11 & 1.0 $\pm$ 0.089 & 0.89 $\pm$ 0.06 & 0.78 $\pm$ 0.14 & 0.26 $\pm$ 0.11 & 1.2 $\pm$ 0.18 \\
        50264    & -0.13$^{10}$ & 0.21$^{10}$ & 0.14$^{10}$ & 1.90$^{10}$ & 0.88$^{11}$ & 1.29$^{11}$ & - - & 1.25$^{11}$ & 1.30$^{10}$ & 1.66$^{10}$ & 1.35$^{10}$ & 0.44$^{11}$ & - - \\
        \hline  
        HR CMa & -0.23 $\pm$ 0.12 & 0.21 $\pm$ 0.071 & 0.25 $\pm$ 0.11 & 0.94 $\pm$ 0.18 & 0.85 $\pm$ 0.18 & 1.0 $\pm$ 0.11 & 0.81 $\pm$ 0.13 & 1.4 $\pm$ 0.13 & 1.0 $\pm$ 0.13 & 0.97 $\pm$ 0.089 & 1.0 $\pm$ 0.14 & 0.31 $\pm$ 0.11 & 1.1 $\pm$ 0.18 \\
            & -0.36$^1$ & 0.25$^1$ & -0.30$^1$ & 1.21$^1$ & 1.08$^1$ & 1.45$^1$ & - - & 1.88$^1$ & 1.56$^1$ & 1.50$^1$ & 1.40$^1$ & 0.54$^1$ & - - \\
        \hline  
        PV UMa & -0.32 $\pm$ 0.13 & -0.13 $\pm$ 0.1 & 0.19 $\pm$ 0.2 & 0.47 $\pm$ 0.15 & 0.14 $\pm$ 0.13 & 0.04 $\pm$ 0.19 & 0.37 $\pm$ 0.19 & 0.62 $\pm$ 0.12 & 0.62 $\pm$ 0.13 & 0.27 $\pm$ 0.14 & 0.47 $\pm$ 0.16 & -0.08 $\pm$ 0.16 & - - \\
            & -0.13$^{2}$ & - - & - - & - - & 0.73$^2$ & 0.75$^2$ & - - & - - & 0.76$^2$ & 0.68$^2$ & - - & - - & - - \\
        \hline          
        HD & -0.35 $\pm$ 0.10 & -0.02 $\pm$ 0.09 & 0.08 $\pm$ 0.16 & 0.53 $\pm$ 0.14 & 0.20 $\pm$ 0.14 & 0.38 $\pm$ 0.13 & 0.56 $\pm$ 0.14 & 0.66 $\pm$ 0.12 & 0.74 $\pm$ 0.12 & 0.65 $\pm$ 0.11 & 0.55 $\pm$ 0.17 & 0.25 $\pm$ 0.11 & 1.1 $\pm$ 0.18 \\
        104979    & -0.26$^{2}$ & 0.03$^{4}$ & -0.23$^{3}$ & 0.99$^{5}$ & 0.71$^{2}$ & 0.85$^{2}$ & - - & 0.52$^{3}$ & 1.11$^{2}$ & 1.06$^{2}$ & 1.13$^{5}$ & 0.40$^{5}$ & - - \\
        \hline  
        HD & -0.38 $\pm$ 0.24 & 0.09 $\pm$ 0.071 & 0.22 $\pm$ 0.2 & 0.29 $\pm$ 0.17 & 0.59 $\pm$ 0.14 & 0.62 $\pm$ 0.14 & 0.64 $\pm$ 0.15 & 1.0 $\pm$ 0.13 & 0.92 $\pm$ 0.14 & 0.69 $\pm$ 0.14 & 0.84 $\pm$ 0.14 & 0.15 $\pm$ 0.18 & - - \\
        31487    & -0.04$^1$ & 0.13$^1$ & - - & - - & - - & 1.11$^1$ & - - & - - & - - & - - & - - & - - & - -\\
        \hline  
        HD & -0.55 $\pm$ 0.29 & 0.4 $\pm$ 0.07 & 0.46 $\pm$ 0.06 & 0.46 $\pm$ 0.11 & 0.35 $\pm$ 0.11 & 0.20 $\pm$ 0.15 & 0.50 $\pm$ 0.18 & 0.42 $\pm$ 0.13 & 0.7 $\pm$ 0.23 & 0.70 $\pm$ 0.15 & 0.52 $\pm$ 0.11 & 0.55 $\pm$ 0.25 & 0.40 $\pm$ 0.22 \\
        101581    & -0.44$^{12}$ & 0.34$^{12}$ & 0.04$^{13}$ & 0.13$^{12}$ & -0.06$^{12}$ & 0.08$^{12}$ & - - & -0.10$^{12}$ & 0.48$^{12}$ & 0.41$^{12}$ & 0.58$^{12}$ & 0.29$^{13}$ & - - \\
        \hline  
        BD & -0.48 $\pm$ 0.18 & 0.37 $\pm$ 0.05 & 0.04 $\pm$ 0.16 & 0.52 $\pm$ 0.15 & 0.60 $\pm$ 0.15 & 0.91 $\pm$ 0.13 & 0.89 $\pm$ 0.13 & 1.40 $\pm$ 0.12 & 1.40 $\pm$ 0.14 & 1.2 $\pm$ 0.10 & 1.20 $\pm$ 0.10 & 0.32 $\pm$ 0.16 & 0.92 $\pm$ 0.18 \\
        +41 2150    & -0.77$^{18}$ & 0.46$^{18}$ & - - & 0.85$^{18}$ & 1.06$^{18}$ & 0.78$^{18}$ & - - & 1.65$^{18}$ & - - & - - & 1.42$^{18}$ & - - & - - \\
        \hline
        CD & -1.35 $\pm$ 0.10 & 0.91 $\pm$ 0.071 & 0.31 $\pm$ 0.11 & 0.54 $\pm$ 0.14 & 0.41 $\pm$ 0.11 & 0.51 $\pm$ 0.16 & $<$ 0.81  & 1.40 $\pm$ 0.13 & 1.00 $\pm$ 0.11 & 1.00 $\pm$ 0.11 & 1.3 $\pm$ 0.18 & 0.51 $\pm$ 0.11 & 2.10 $\pm$ 0.14 \\
        -62 1346 & -1.59$^{24}$ & 0.86$^{24}$ & 0.69$^{24}$ & - - & 0.46$^{24}$ & 0.86$^{24}$ & - - & 1.58$^{24}$ & 1.18$^{24}$ & 1.24$^{24}$ & 1.25$^{24}$ & - - & 2.05$^{24}$ \\
        \hline
        HD & -1.41 $\pm$ 0.24 & 0.61 $\pm$ 0.11 & 0.27 $\pm$ 0.06 & 0.37 $\pm$ 0.26 & -0.06 $\pm$ 0.15 & 0.04 $\pm$ 0.16 & 0.24 $\pm$ 0.26 & 0.14 $\pm$ 0.13 & 0.12 $\pm$ 0.16 & 0.07 $\pm$ 0.11 & 0.14 $\pm$ 0.14 & 0.14 $\pm$ 0.18 & - - \\
        135148    & -2.17$^6$ & 0.80$^7$ & - - & - - & - - & - - & - - & 0.30$^7$ & 0.31$^9$ & - - & - - & 0.71$^9$ & - - \\
        \hline  
        HD & -1.60 $\pm$ 0.40 & 1.50 $\pm$ 0.21 & 0.76 $\pm$ 0.20 & 1.10 $\pm$ 0.23 & 0.95 $\pm$ 0.25 & 1.5 $\pm$ 0.11 & $<$ 1.30 & 1.50 $\pm$ 0.11 & $<$ 1.80 & 1.90 $\pm$ 0.18 & 1.90 $\pm$ 0.20 & 1.10 $\pm$ 0.20 & 2.06 $\pm$ 0.22 \\
        209621    & -1.99$^{15}$ & 1.26$^{15}$ & 0.17$^{17}$ & 1.02$^{16}$ & 0.36$^{16}$ & 1.80$^{16}$ & - - & 1.76$^{15}$ & 2.41$^{16}$ & 2.04$^{16}$ & 1.87$^{16}$ & 1.35$^{16}$ & 1.88$^{16}$ \\
        \hline  
        TYC & -1.80 $\pm$ 0.25 & 0.61 $\pm$ 0.11 & 0.22 $\pm$ 0.16 & -0.02 $\pm$ 0.21 & -0.06 $\pm$ 0.13 & $<$ 0.61 & $<$ 0.81 & 0.76 $\pm$ 0.12 & 0.56 $\pm$ 0.16 & 0.71 $\pm$ 0.14 & $<$ 0.41 & $<$ 0.36 & 1.4 $\pm$ 0.27 \\
        1987-753-1    & -2.05$^{15}$ & 2.39$^{15}$ & - - & - - & - - & - - & - - & 0.75$^{15}$ & - - & - - & - - & 0.04$^{15}$ & - - \\
        \hline 
        HE & -1.89 $\pm$ 0.15 & 0.95 $\pm$ 0.07 & 0.35 $\pm$ 0.14 & 0.78 $\pm$ 0.2 & 0.65 $\pm$ 0.25 & 1.00 $\pm$ 0.16 & 0.59 $\pm$ 0.26 & 1.40 $\pm$ 0.13 & 1.30 $\pm$ 0.15 & 1.30 $\pm$ 0.16 & 1.30 $\pm$ 0.13 & 0.85 $\pm$ 0.11 & 2.20 $\pm$ 0.18 \\
        0414-0343    & -2.50$^{23}$ & 1.3$^{23}$ & 0.49$^{22}$ & 0.48$^{22}$ & 0.20$^{22}$ & 0.52$^{22}$ & - - & 1.9$^{23}$ & 1.48$^{22}$ & 1.42$^{22}$ & 1.63$^{22}$ & 1.23$^{22}$ & 2.53$^{22}$ \\
        \hline  
        BD & -1.93 $\pm$ 0.30 & 0.90 $\pm$ 0.10 & 0.33 $\pm$ 0.05 & 0.36 $\pm$ 0.14 & 0.27 $\pm$ 0.11 & 0.60 $\pm$ 0.17 & 0.70 $\pm$ 0.22 & 1.2 $\pm$ 0.11 & 1.10 $\pm$ 0.089 & 1.10 $\pm$ 0.11 & 0.91 $\pm$ 0.098 & 0.50 $\pm$ 0.16 & 1.70 $\pm$ 0.18 \\
        +04 2466    & -1.90$^{15}$ & 1.24$^{15}$ & 0.24$^{17}$ & - - & 0.49$^{19}$ & - - & - - & 1.63$^{15}$ & 1.31$^{19}$ & - - & 1.23$^{19}$ & 0.37$^{19}$ & 1.92$^{17}$ \\
        \hline  
        HD & -2.02 $\pm$ 0.17 & 0.47 $\pm$ 0.07 & 0.44 $\pm$ 0.11 & 0.30 $\pm$ 0.26 & 0.12 $\pm$ 0.19 & 0.44 $\pm$ 0.11 & $<$ 1.10 & -0.17 $\pm$ 0.16 & 0.17 $\pm$ 0.09 & 0.21 $\pm$ 0.16 & -0.13 $\pm$ 0.16 & 0.37 $\pm$ 0.22 & $<$ -0.33 \\
        103545    & -2.45$^9$ & -0.04$^9$ & - - & - - & -0.26$^{14}$ & - - & - - & - - & 0.17$^{14}$ & - - & - - & 0.37$^{14}$ & 0.65$^{14}$ \\
        \hline  
        HD & -0.04 $\pm$ 0.17 & 0.00 $\pm$ 0.19 & 0.20 $\pm$ 0.11 & 0.12 $\pm$ 0.14 & -0.26 $\pm$ 0.10 & 0.20 $\pm$ 0.11 & $<$ 0.50 & -0.23 $\pm$ 0.13 & 0.30 $\pm$ 0.20 & -0.20 $\pm$ 0.19 & $<$ 0.30 & -0.20 $\pm$ 0.20 & $<$ 0.90 \\
        116514    & -0.15$^{21}$ & - - & - - & - - & - - & - - & - - & - - & - - & - - & - - & - - & - - \\
        \hline  
        TYC & -0.08 $\pm$ 0.08 & -0.24 $\pm$ 0.11 & 0.01 $\pm$ 0.16 & 0.21 $\pm$ 0.15 & -0.02 $\pm$ 0.11 & 0.14 $\pm$ 0.13 & 0.31 $\pm$ 0.2 & 0.29 $\pm$ 0.12 & 0.26 $\pm$ 0.14 & 0.26 $\pm$ 0.14 & 0.36 $\pm$ 0.14 & 0.21 $\pm$ 0.16 & $<$ 0.41 \\
        8258-1189-1 & -0.14$^{25}$ & -0.05$^{25}$ & 0.01$^{25}$ & - - & 0.14$^{25}$ & -0.03$^{25}$ & 1.37$^{25}$ & 0.66$^{25}$ & 0.00$^{25}$ & -0.10$^{25}$ & - - & 0.09$^{25}$ & - - \\
        \hline
        TYC & -0.31 $\pm$ 0.19 & 0.28 $\pm$ 0.094 & 0.31 $\pm$ 0.16 & 0.71 $\pm$ 0.10 & 0.38 $\pm$ 0.13 & 0.40 $\pm$ 0.11 & 0.74 $\pm$ 0.18 & 0.98 $\pm$ 0.12 & 0.96 $\pm$ 0.10 & 0.83 $\pm$ 0.10 & 0.93 $\pm$ 0.11 & 0.38 $\pm$ 0.11 & 1.2 $\pm$ 0.18 \\
        2250-1047-1    & -0.55$^{20}$ & 0.37$^{20}$ & - - & 1.07$^{20}$ & 0.44$^{20}$ & 0.77$^{20}$ & - - & 1.17$^{20}$ & 1.18$^{20}$ & 0.97$^{20}$ & - - & - - & - -\\
        \hline  
        HD & -0.31 $\pm$ 0.12 & 0.11 $\pm$ 0.21 & 0.08 $\pm$ 0.09 & -0.14 $\pm$ 0.17 & -0.24 $\pm$ 0.12 & 0.16 $\pm$ 0.25 & $<$ 0.46 & -0.13 $\pm$ 0.2 & 0.26 $\pm$ 0.14 & $<$ -0.32 & 0.29 $\pm$ 0.25 & $<$ 0.46 & - - \\
        51273    & -0.80$^{21}$ & - - & - - & - - & - - & - - & - - & - - & - - & - - & - - & - - & - - \\
        \hline  
        HD & -0.35 $\pm$ 0.18 & -0.51 $\pm$ 0.16 & 0.47 $\pm$ 0.045 & -0.18 $\pm$ 0.18 & -0.45 $\pm$ 0.18 & -0.12 $\pm$ 0.19 & 0.25 $\pm$ 0.18 & -0.36 $\pm$ 0.13 & -0.05 $\pm$ 0.16 & -0.37 $\pm$ 0.08 & -0.1 $\pm$ 0.2 & -0.05 $\pm$ 0.25 & - - \\
        33363    & -0.05$^{21}$ & - - & - - & - - & - - & - - & - - & - - & - - & - - & - - & - - & - - \\
        \hline 
        TYC & -0.51 $\pm$ 0.13 & -0.02 $\pm$ 0.11 & 0.18 $\pm$ 0.12 & 0.1 $\pm$ 0.16 & 0.19 $\pm$ 0.16 & 0.53 $\pm$ 0.16 & 0.53 $\pm$ 0.17 & 1.0 $\pm$ 0.12 & 0.91 $\pm$ 0.13 & 0.93 $\pm$ 0.13 & 1.1 $\pm$ 0.14 & 0.33 $\pm$ 0.14 & $<$ 1.0 \\
        2866-338-1    & - - & - - & - - & - - & - - & - - & - - & - - & - - & - - & - - & - - & - - \\
        \hline  
        BD & -0.67 $\pm$ 0.13 & -0.06 $\pm$ 0.16 & 0.44 $\pm$ 0.16 & 0.17 $\pm$ 0.17 & -0.18 $\pm$ 0.15 & 0.31 $\pm$ 0.13 & 0.34 $\pm$ 0.18 & 0.04 $\pm$ 0.13 & 0.14 $\pm$ 0.13 & 0.04 $\pm$ 0.15 & 0.27 $\pm$ 0.13 & 0.34 $\pm$ 0.11 & 0.14 $\pm$ 0.14 \\
        -19 3868 & -0.63$^{26}$ & - - & 0.27$^{26}$ & - - & - - & - - & - - & - - & - - & - - & - - & - - & - - \\
        \hline
        HD & -0.90 $\pm$ 0.18 & 0.72 $\pm$ 0.16 & -0.05 $\pm$ 0.22 & 0.63 $\pm$ 0.12 & 0.53 $\pm$ 0.14 & $<$ 0.66 & $<$ 0.73 & 0.76 $\pm$ 0.22 & 0.96 $\pm$ 0.26 & 0.88 $\pm$ 0.2 & $<$ 1.1 & $<$ 0.93 & - - \\
        276679    & - - & - - & - - & - - & - - & - - & - - & - - & - - & - - & - - & - - & - - \\
        \hline\hline
    \end{tabular}
    \label{tab:abund_compare}
    \tablefoot{Upper limits are noted with $<$. For the stars HD 33363, HD 51273, HD 276679, HD 116514, and TYC 2866-338-1 we were unable to find abundance information in the literature.}
    \tablebib{(1)\cite{2018A&A...618A..32K} (2)\cite{2019A&A...626A.127J} (3)\cite{2015A&A...574A..50J} (4)\cite{2016ApJ...833..181C} (5)\cite{2015MNRAS.446.2348K} (6)\cite{2014ApJ...797...21P} (7)\cite{2011MNRAS.412..843S} (8)\cite{2018A&A...615A..76S} (9)\cite{2004ApJ...617.1091S} (10)\cite{2019MNRAS.486.3266P} (11)\cite{2003A&A...402.1061P} (12)\cite{2018AJ....155..111L} (13)\cite{2017A&A...606A..94D} (14)\cite{2010ApJ...724..975R} (15)\cite{2016ApJ...833...20Y} (16)\cite{2010MNRAS.404..253G} (17)\cite{2011MNRAS.418..284B} (18)\cite{2009A&A...508..909Z} (19)\cite{2013ApJ...771...67I} (20)\cite{2021A&A...654A.140K} (21)\cite{2022yCat.1355....0G} (22)\cite{2015ApJ...814..121H} (23)\cite{2016A&A...588A..37H} (24)\cite{2012A&A...543A..58P}, (25)\cite{2021MNRAS.506..150B}, (26)\cite{2020AJ....160...82S}}
\end{sidewaystable*}

\FloatBarrier

\section{Stellar orbital parameters}

\begin{sidewaystable*}[t]
    \centering
    \caption{Orbital parameters and estimated stellar masses where sufficient data exists to model the orbits using ELC.}
    \small
    \begin{tabular}{l c r r r r r r r r r c}
    \hline \hline
    Star & \# Obs. &  $\omega$ & $e$ & $P$ & $k_1$ & $T_0$      & $a\sin{i}$ &    $f(m)$    &    $M_1$   &     $M_2$    & Ref. \\
         & \# Lit. & ($^{\circ}$) &  & (d) & (km/s)&(JD-2400000) & ($10^9$m) & (M$_{\odot}$) & (M$_{\odot}$) & (M$_{\odot}$) & \\

    \hline \hline
    HD 105671       &   1 & 244.46 $\pm$       & 0.138 $\pm$ 0.050 &   817.60 $\pm$  95.75 &  0.75 & 60605.361 $\pm$ 284.127 &   111.94 & 3.2e-5 & 5.33 $\pm$ 0.75 & 0.34 $\pm$ 0.25 & \\
                    &   6 & - -                &  - -              &     - -               &  - -  & - -                     &   -  -  &  - -   & 7.40            & - -  & (6,7) \\
    \hline
    CD-62 1346      &   1 & 355.93 $\pm$ 11.97 & 0.340 $\pm$ 0.060 &   358.18 $\pm$   2.72 &  2.31 & 59938.048 $\pm$  15.724 &    70.76 & 3.9e-4 & 2.89 $\pm$ 0.87 & 0.50 $\pm$ 0.23 & \\
                    &  12 & - -                & - -               & - -             & - - &   - - & - -                     &  - -    & - -   & - -             & - - \\
    \hline
    HD 50264        &   5 & 249.25 $\pm$  8.37 & 0.011 $\pm$ 0.006 &   912.40 $\pm$   0.38 &  9.52 & 49579.283 $\pm$. 20.442 &   271.76 & 0.069 & 1.25 $\pm$ 0.89 & 0.66 $\pm$ 0.32 & \\
                    &  11 & 230                & 0.091             &   909.90              &  9.52 & 49541                   &   -  -  &  - -  & 0.90            & 0.60 & (6) \\
    \hline
    HD 196673       &   1 & 116.19 $\pm$  1.49 & 0.589 $\pm$ 0.016 &  7784.88 $\pm$  75.12 &  3.58 & 51697.923 $\pm$  77.760 &  1823.0 & 0.019 & 5.08 $\pm$ 0.32 & 1.05 $\pm$ 0.04 & \\
                    &  79 & 116                & 0.590             &  7780.00              &  3.70 & 51698                   &   314.0 & 0.020 & 5.00            & 1.10 & (4,6) \\
    \hline
    HD 104979       &   2 & 164.34 $\pm$  9.59 & 0.075 $\pm$ 0.022 & 18189.37 $\pm$ 621.76 &  1.65 & 60297.864 $\pm$ 149.790 &  1466.92 & 0.008 & 2.99 $\pm$ 0.65 & 1.10 $\pm$ 0.34 & \\
                    & 128 & - -                & 0.080             & 19295.02              &  - -  & - -                     &   -  -  &  - -  & 2.70            & 0.75 & (6) \\
    \hline
    HR CMa          &   6 & 180.80*            & 0.055 $\pm$ 0.020 &   457.32*             &  6.79 & 47357.722 $\pm$  55.167 &  180.13 & 0.014 & 1.87 $\pm$ 0.95 & 0.60 $\pm$ 0.15 & \\
                    &  38 & - -                & 0.013             &   457.40              &  - -  & - -                     &   -  -  & 0.035 & 2.10            & 0.75 & (5,6,10) \\
    \hline
    PV UMa          &   3 &  22.69 $\pm$  2.10 & 0.110 $\pm$ 0.004 &    80.52 $\pm$  0.001 &  8.48 & 55155.728 $\pm$   0.061 &    32.48 & 0.005 & 2.20 $\pm$ 0.64 & 1.01 $\pm$ 0.10 & \\
                    & 140 & - -                & 0.090             &    80.00              & - -   & - -                     &   -  -  & 0.005 & 3.90            & 0.94 & (6,8,9) \\
    \hline
    HD 31487        &   1 & 246.02 $\pm$ 60.45 & 0.049 $\pm$ 0.008 &  1086.80 $\pm$  72.23 &  7.86 & 49432.182 $\pm$ 123.928 &   255.57 & 0.060 & 1.95 $\pm$ 0.98 & 0.73 $\pm$ 0.33 & \\
                    &  36 & 238                & 0.050             &  1066.40              &  7.00 & 45173                   &   -  -  & 0.038 & 3.40            & 1.59 & (4,5) \\
    \hline
    BD+41 2150      &   3 & 358.86 $\pm$  1.60 & 0.057 $\pm$ 0.001 &   306.82 $\pm$   2.04 & 10.78 & 56323.099 $\pm$   1.822 &   187.69 & 0.040 & 1.69 $\pm$ 0.94 & 0.88 $\pm$ 0.31 & \\
                    &  70 & 315                & 0.055             &   323.0               & 11.80 & 56114                   &   -  -  &  - -  & 1.27            & - -  & (8,12,13,14,15) \\
    \hline
    HD 135148       &  11 & 261.04 $\pm$ 18.45 & 0.114 $\pm$ 0.020 &  1405.19 $\pm$   2.78 &  4.87 & 47477.107 $\pm$  67.530 &   496.36 & 0.016 & 2.72 $\pm$ 1.73 & 0.72 $\pm$ & 0.35 \\
                    &  36 & 273                & 0.123             &  1416                 &  4.79 & 47534                   &    -  - &  - -  & 0.80            & 0.26 & (1) \\    
    \hline
    HD 209621       &   1 &  15.66 $\pm$ 10.55 & 0.005 $\pm$ 0.005 &   407.81 $\pm$   0.42 & 11.98 & 45877.766 $\pm$  15.605 &   247.32 & 0.072 & 2.30 $\pm$ 1.10 & 1.15 $\pm$ 0.24 & \\
                    &  27 & - -                & 0.000             &   407.4               & 12.05 & 45858                   &    -  - &  - -  & 3.60            & - -  & (4,8,11,12,13) \\
    \hline
    BD+04 2466      &   2 & 319.91 $\pm$  2.23 & 0.166 $\pm$ 0.022 &  4762.26 $\pm$  20.12 &  5.53 & 59908.329 $\pm$  12.918 &  1131.9 & 0.080 & 2.40 $\pm$ 1.20 & 1.18 $\pm$ & 0.21 \\
                    &  42 & - -                &  - -              &  4600                 &  - -  & - -                     &   -  -  &  - -  & - -             & 1.10 & (11,16) \\ 
    \hline \hline
    \end{tabular}
    \tablefoot{Uncertainties in the parameters are taken from the widths of the posterior distributions after MCMC optimization. For each star, the first line are the results from our analysis and the second line are the data from the literature.}
    \tablebib{(1)\cite{2003AJ....125..293C} (2)\cite{2008MNRAS.389.1722E} (3)\cite{1989AJ.....97.1139F} (4)\cite{1990ApJ...352..709M} (5)\cite{2017A&A...597A..68V} (6)\cite{2019A&A...626A.127J} (7)\cite{2019A&A...624A..78D} (8)\cite{2017A&A...608A.100E} (9)\cite{2013AJ....145...41K} (10)\cite{1998A&AS..131...25U} (11)\cite{2011MNRAS.418..284B} (12)\cite{2016A&A...586A.158J} (13)\cite{2020A&A...639A..24E} (14)\cite{2003A&A...397..997P} (15)\cite{2016ApJ...826...85S} (16)\cite{2015A&A...581A..22A} (17)\cite{1999A&AS..137..369F} (18)\cite{1991JApA...12..289G}}
    \label{tab:orbit_params_out}
\end{sidewaystable*}

\FloatBarrier

\section{Discussion on individual stars}
\subsection{Ba stars}

\paragraph{HD 105671} is metal rich, with [Fe/H] = 0.10, and was classified as mild barium dwarf by \cite{2019A&A...626A.127J}. Using Xiru, we estimate atmospheric parameters and our results are compatible with those determined by \emph{Gaia}, both photometrically and spectroscopically. Our abundances are in good agreement with \cite{2019A&A...624A..78D} and \cite{2018A&A...615A..76S}, and we add Mo, La, Nd, Eu, and Pb to the pattern. Adding these elements helps constrain the mass of the AGB star that produced the s-process elements, and provides a constraint on our orbital modelling. This Ba star shows near-solar abundances in both s-process peaks with some scatter. We add one RV data point to the existing six in the literature. However, we were not able to detect appreciable variability in the RVs and constrain the orbit. Follow up observations will provide more RV data to characterize the binarity of this system. 

\paragraph{HD 196673} is a mild Ba dwarf \citep{1990ApJ...352..709M} recently studied by \citet{2019A&A...626A.127J}, who derived a metallicity of [Fe/H] = -0.12, closely matching our derived metallicity using Xiru of [Fe/H] = 0.00 $\pm$ 0.14. Effective temperatures derived using Xiru are also in good agreement with \citet{2019A&A...626A.127J}, although our estimated surface gravity is smaller at $0.94$ compared to $2.40$. As a result, we compute abundances that are lower than those determined by \citet{2019A&A...626A.127J}. Abundances in this star are closer to solar values compared to strong Ba stars in our sub-sample, and result in a relatively flat profile in Figure \ref{fig:FRUITY_compare} with a dilution factor $\delta = 0.4$, signifying mixing in the companion. This type of profile is indicative of a mild Ba star, like HD 105671; the near-solar abundances could be a side effect of a comparatively noisier spectrum with lower S/N $\sim30$. Our derived orbital and physical parameters are in good agreement with those from \citet{1990ApJ...352..709M} and \citet{2019A&A...626A.127J}, with a high eccentricity $e\sim0.6$ and a long orbital period of $\sim7646$ days. The visible component has a relatively high mass of $\sim5 M_{\odot}$, and the unseen component has a mass within the Chandrasekhar limit, confirming the likelihood of being a white dwarf left from an exhausted AGB star. This star may have accreted material from its previous AGB companion only during times around periastron, resulting in the mild abundance pattern.

\paragraph{HD 50264} is a CH sub-giant star \citet{2003A&A...402.1061P} with strong enhancements in s-process material ([s/Fe] = 0.83). Stellar parameters estimated using Xiru are in good agreement with those in \citet{2019MNRAS.486.3266P}, with any differences being covered by the relative uncertainties. We observe a strong s-process signature in the computed abundance pattern. Our comparison to the FRUITY yields suggests an AGB mass of 1.50 $M_{\odot}$ and a moderate dilution factor, meaning the AGB material has been somewhat mixed. This star exists in a nearly circular ($e = 0$) long period ($\sim 900$ days) binary, and is a prototypical example of a metal-rich ([Fe/H] = -0.17 $\pm$ 0.07) star that has been polluted by an AGB companion that has since become a white dwarf.

\paragraph{HR CMa} is a well-studied eclipsing Ba giant. With significant enhancements in s-process material ([s/Fe] = 0.91) and confirmed binarity, this is a de-facto Ba star. Atmospheric parameters derived through ionization and excitation balance are similar to that from the APOGEE survey infrared SED fitting, although our temperature is about $100$ K higher. Our abundances are slightly lower than those measured by \cite{2018A&A...618A..32K} by about 0.40 dex. We add Mo and Pb to the pattern, and we find a good fit to the surface abundances with an AGB of initial mass 2.5 M$_{\odot}$ and a dilution factor of 0.4 (Eq. \ref{eqn:dilute}; Fig. \ref{fig:FRUITY_compare}), indicating significant mixing. We add a few RV data points to further constrain the orbit. We determine orbital parameters $P$ and $e$ in agreement with the literature, while our derived masses are slightly lower than previously estimated values. These are, however, still compatible with the AGB accretion scenario for the observed s-process enhancement.

\paragraph{PV UMa} is a short-period, mild Ba star binary with mild enhancements in s-process material ([s/Fe] = 0.37). We estimate the metallicity using Xiru for [Fe/H] = -0.32 $\pm$ 0.13, slightly lower than previous estimates, but just outside of the 1$\sigma$ uncertainties. Our temperature of 5058 K is in excellent agreement with the 5050 K from \cite{2019A&A...626A.127J}, and our surface gravity estimate from Xiru of $2.28$ compares nicely to the $2.5$ from \cite{2019A&A...626A.127J}. We find lower abundances in the first and second s-peak elements, and we add Sr, Mo, Nd, and Eu to the abundance pattern. The orbit of PV UMa is lightly eccentric ($e=0.10$) with a period of only $80$ days, quite short for a Ba star to have received its material via AGB wind mass transfer. This system may have accreted material via RLOF, and undergone common envelope evolution with the previous AGB star, resulting in the less-pronounced mild s-process pattern.

\paragraph{HD 104979} is a mild Ba giant with enhancements in s-process elements ([s/Fe] $\sim$ 0.60). Our derived atmospheric parameters from Xiru are in good agreement with the APOGEE survey results. \citet{2019A&A...626A.127J} found larger enhancements in Y, Zr, La, and Ce than this study by about $0.4$ dex. These differences can partially be attributed to the differences in computed metallicity. In our abundance analysis, we add the elements Mo and Pb. Our abundance analysis suggests enhancement from a more massive AGB star, around $4.0 M_{\odot}$. \citet{2019A&A...626A.127J} investigated the binarity of this system, with a nearly circular $e=0.07$ orbit and a long period of $P=18370$ days. From our orbital analysis, we derive similar orbital parameters $e=0.08, P=19295$, and estimate component masses $M_1=1.65, M_2=0.75$, where the mass of the unseen companion is in agreement with \citet{2019A&A...626A.127J}. 

\paragraph{HD 31487} is a Ba star with significant enhancements in heavy elements, and we add Sr, Y, Mo, Ba, La, Ce, Nd, and Eu to the abundance pattern. Our estimate of the metallicity using Xiru is in relative agreement with \citet{2018A&A...618A..32K}, within combined uncertainties in the two estimates, as well as the effective temperature and surface gravity. Our computed Zr abundance is lower than \citet{2018A&A...618A..32K} by about $0.4$ dex, and this discrepancy can be directly attributed to the difference in metallicity. Through their orbital analysis, \citet{2023A&A...671A..97E} found a high white dwarf mass of $1.59\pm0.22$ M$_{\odot}$ corresponding to a barium star mass of $3.4\pm0.2$. In our optimization, we determine a primary mass of $1.97\pm$ M$_{\odot}$, more consistent with the average barium star mass and and close to the $1.5$ M$_{\odot}$ model in our abundance analysis, and a white dwarf mass of $0.73\pm$ M$_{\odot}$m, which is within the Chandrasekhar mass limit for white dwarfs. 

\paragraph{HD 101581} is a high proper motion star, and a Ba dwarf \citep{2019A&A...626A.127J} with metallicity ([Fe/H] = -0.55 $\pm$ 0.29). Parameters derived to characterize the atmosphere of this star with Xiru are slightly discrepant from those provided by \emph{Gaia}, where our estimated $T_{eff}$ is higher, our surface gravity agrees within the error bar, and the metallicity is in good agreement. Our abundance analysis shows enhancements in magnesium and and we find larger enhancements in s-process elements ([s/Fe] = 0.47), compared with \citet{2018AJ....155..111L}; see Table \ref{tab:abund_compare}. We add Pb to the abundance pattern, and constrain the initial AGB mass to about 1.3 M$_{\odot}$, with significant mixing ($\delta = 0.3$). We conclude this is a mild Ba star with a less-pronounced light s-element peak, similar to PV UMa. \citet{2019A&A...626A.127J} did not detect variability in the RVs, but more RV data points will reveal whether or not this star exists in an binary. Our orbital analysis points towards a long period ($\approx 750$ days) in an eccentric orbit ($e \approx 0.24$), but we find these results are not definitive. This star will continue to be observed in our monitoring program to provide more constraints on the binary orbit.

\subsection{CH stars}

\paragraph{BD+41 2150} is a CH giant in a spectroscopic binary, as confirmed by \cite{2003A&A...397..997P}. We find our atmospheric parameters in agreement with the analysis of the \emph{Gaia} BP/RP spectra by \citet{2023MNRAS.524.1855Z}. This star is enhanced in s-process material, as shown by \cite{2009A&A...508..909Z} who measured an enhancement of [s/Fe]=1.26. We find comparable enhancement, with [s/Fe]=1.00. \citet{2009A&A...508..909Z} concluded this star has been polluted by an AGB companion, based on its abundance pattern, and our analysis finds comparable results. We find a good match in the abundances compared to a model of a 2.0 M$_{\odot}$ progenitor AGB star with a dilution factor of $\delta=0.3$, indicative of significant mixing. Through orbital modelling, we find the period of the binary to be just over $300$ days with an orbital separation of about $1.1$ AU. Such a small orbit may be indicative of other mass transfer scenarios, as opposed to only AGB wind mass-transfer. Our estimated mass of the visible CH star is $1.69$ M$_{\odot}$. It is likely that the CH star in the system will share mass with its companion as it evolves up the RGB and onto the AGB, polluting the now white dwarf with heavy elements. \citet{2020A&A...639A..24E} discusses the orbital period and location in the HR diagram, and we agree that this is an RGB star that will likely go into a Roche-lobe overflow phase before helium burning. 

\paragraph{CD-62 1346} is an evolved metal-poor giant star, and our atmospheric parameters are in good agreement with those from \citet{2012A&A...543A..58P}, despite a slightly warmer surface temperature predicted by Xiru. A high proper-motion halo star, \citet{2012A&A...543A..58P} suggest using similar stars to constrain the Galactic potential. A large Pb to Ce ratio means this is another ``lead'' star, and in our abundance analysis we find abundances in very good agreement, with [Pb/Ce] $> 0.80$. We add Sr, Mo, and Eu to the abundance pattern. This star shows significant RV variation \citep{2019A&A...626A.128E}, and we contribute a small number of RV data points to the orbit. We perform an orbital analysis with wide prior distribution on orbital and physical stellar parameters. Our results indeed confirm the binary nature of this star, and we estimate an orbital period of $357$ days with an eccentricity of $e=0.35$. This is the first time the orbit for CD-62 1346 has been computed, but we note the large uncertainties on our orbital parameter estimates in Table \ref{tab:orbit_params_out}. Upcoming follow-up RV observations will help refine these orbital parameters.

\paragraph{HD 209621} is a well-studied metal-poor CH star, also classified as CEMP-r/s star \citep{2001ApJ...557..802B}. Xiru derived stellar parameters are similar to those by \cite{2021A&A...654A.140K}, with a slightly higher temperature and higher metallicity. Our abundances are in good agreement with those found by \citet{2010MNRAS.404..253G}, and we add Mo to the abundance pattern. We find [Zr/Y] $approx$ 0.75, in support of the large [Zr/Y] ratio found by \citet{2012MNRAS.422..849B} where only one Zr feature was observed; we confirm this result with multiple Zr and Y lines. From our orbital analysis, we find this star to be in a nearly circular long period binary with an orbit of just over $400$ days, supporting the AGB mass transfer hypothesis. We find a good fit to a 2.0 $M_{\odot}$ AGB yield with moderate mixing ($\delta = 0.4$) in our comparison to the FRUITY database.

\paragraph{BD+04 2466} is a metal-poor ([Fe/H] = -1.93 $\pm$ 0.30 CH star \citep{2009A&A...496..791P} that shows large abundances of heavy elements ([s/Fe] $\sim$ 0.89) and displays characteristics of the mass-transfer paradigm. Our parameters estimated using Xiru are in good agreement with \emph{Gaia} photometry and GALAH spectroscopic parameters. BD+04 2466 is a known `Pb star' due to its high [Pb/Ce] ratio. We find good agreement in our abundance analysis compared to \citet{2011MNRAS.418..284B}, who also detected large enhancements in s-process material and low enhancements in r-process elements like Eu. We add the elements Sr, Zr, and Mo to the abundance pattern. We find a good fit to an initial AGB mass of 2.0 M$_{\odot}$ with a dilution factor of 0.2, indicating significant mixing of the AGB material. \citet{2005A&A...441.1135J} found this star to be in a wide binary system with a long period of about $4600$ days; in our orbital optimization we find compatible results (\ref{tab:orbit_params_out}, giving further support to the mass-transfer hypothesis. From our orbital modelling, we estimate a mass of $2.4$ solar masses for the visible component. There is good agreement between the FRUITY models and our ELC dynamical estimate of the mass. Our secondary mass is slightly larger than expected from an initial 2.0 M$_{\odot}$ mass AGB star, with a white dwarf mass of $1.1$ solar masses. 

\subsection{CEMP (-s/-no) stars}

\paragraph{HD 135148} is a CEMP-no/r star that shows small enhancements in s-process material ([s/Fe] $\sim 0.22$). We find a slightly higher metallicity using Xiru ([Fe/H] = $-1.41$) compared to previous studies; this result could be due to strong NLTE effects in the ionization and excitation balance of the Fe II lines in more metal-poor stars. Considering differences in atmospheric parameters, our abundances are in decent agreement with \citet{2004ApJ...617.1091S}. This star is a known binary system with a long period, but the low s-process enhancement relative to the r-process element Eu signals star is not likely to have been enriched by a companion AGB star via mass transfer. This is reflected in the flat fit to the FRUITY models and a very high AGB mass of $6.0 M_{\odot}$.

\paragraph{TYC 1987-753-1} is a CEMP-s star \citep{2016ApJ...833...20Y}, and we add 8 elements (Sr, Y, Zr, Mo, La, Ce, Nd, and Pb) to the abundance pattern first described by \cite{2012A&A...548A..34A}. Using Xiru, we find similar atmospheric parameters compared to the recent study by \citet{2021ApJ...907...10L}. The low Eu abundance suggests no significant enhancements in r-process material, supporting the CEMP-s classification with little contributions from the r-process, even from the parent molecular cloud. The abundances in the light s-process peak are somewhat scattered, but the second peak and Pb are fit nicely by the FRUITY yields with an initial AGB mass of 2.0 $M_{\odot}$ and a dilution factor of $0.1$, descriptive of significant mixing within the observed star. We find discrepancies in RV between our data and the \emph{Gaia} DR2 and DR3 radial velocities, hinting at a binary system, but we currently lack sufficient time-series RV data to perform a full characterization of the possible binary orbit.

\paragraph{HE 0414-0343} was determined to be a CEMP-s star by \citet{2006ApJ...652.1585F}, and \citet{2016A&A...588A..37H} found large enhancements in C, N, Sr, and Ba, who also note the possibility of HE 0414-0343 being an AGB star. From Xiru, our atmospheric parameters are not in full agreement with previous studies. The temperature estimated by Xiru is only $200$ K higher than previous studies, but our surface gravity and metallicity are in decent agreement with photometric results from \emph{Gaia} DR3. Our surface gravity from Xiru $\log g = 2.13 \pm 0.62$ could hint towards an AGB star, provided the higher-than-usual uncertainties, but we do not find this conclusive. We were unable to detect Tc lines in absorption within the high S/N, high resolution spectrum, indicating that this star is not actively producing its own heavy metals, but received its s-process material from a former AGB companion. \citet[][]{2015ApJ...814..121H} found a low surface gravity of $\log g = 1.25$, and reports high carbon and s-process enhancements. Our carbon enhancement of [C/Fe] = $0.95$ is significant, and in agreement with \citet{2016A&A...588A..37H}. Our high-resolution observations confirm the large enhancements in s-process material, and we add Mo to the pattern. While our first s-peak element abundances are in relative agreement with previous studies, the second s-peak elements are slightly under-abundant compared to \citet{2016A&A...588A..37H}. Our FRUITY yield fits indicate an initial AGB mass of 1.5 M$_{\odot}$ with substantial mixing ($\delta = 0.1$). This is in decent agreement with the 1.3 M$_{\odot}$ estimation of \citet{2013ApJ...770..104P} and \citet{2015ApJ...814..121H}. This star is a known spectroscopic binary exhibiting significant variations in the observed RVs, indicating an unseen binary companion \citep{2015ApJ...814..121H}. However, with only a few data points, we were unable to fully model the orbit of the binary. 

\paragraph{HD 103545} is a CEMP-no ([Fe/H]=-1.83 $\pm$ 0.17) star with lightly scattered heavy element abundance patterns \citep{2004ApJ...617.1091S}. This is a case where Xiru performed well on a metal-poor, carbon-enhanced star, where our atmospheric parameters are in good agreement to the APOGEE values, described in \cite{2020AJ....160..120J} and those in \citet{2021ApJ...907...10L}. Similarly to \citet{2014ApJ...797...21P}, we do not find strong enhancements in s-process elements for this star ([s/Fe] = -0.12). The FRUITY models do not suggest that the surface abundance pattern is similar to that of an AGB star, and provide a dilution factor of $\delta = 0.1$ with an initial AGB mass of 6.0 M$_{\odot}$, similar to our analysis of HD 135148. According to \citet{2014ApJ...797...21P}, HD 103545 is a CEMP-no star with a [Ba/Fe] < 0, a [Eu/Fe] = 0.46, and $\log\epsilon$(C)=6.44 \citep{2013A&A...552A.107S}. This star and HD 135148 have similarly flat abundance patterns, both separating from the rest of the C-enhanced stars in the top panel of Figure \ref{fig:abund_pattern}. We do not detect significant RV variation in the system with only a few RV data points, moving further away from any AGB mass-transfer hypothesis.

\subsection{``Other'' stars}\label{sec:apx_other}

\paragraph{HD 116514} is a G type main-sequence star, and we observe a small enhancement in s-process material ([s/Fe] = $+0.18$). Atmospheric parameters determined by Xiru include a slightly higher $T_{eff}$, $\sim$6250 compared to $\sim$5800 from \emph{Gaia} and APOGEE, where the $\log g$ and [Fe/H] are in good agreement. HD 116514 displays a weak s-process profile in our comparison to the FRUITY models, similar to that of the mild Ba stars HD 105671 and HD 196673. The low dilution factor indicates that any AGB material that has been accreted has been significantly mixed. Our orbital analysis shows this system is in a close orbit with a period of only $5$ days. This is too close for wind mass transfer from an AGB companion where typical orbital periods are on the order of 1000 days. However, this may be a result of post common envelope evolution, further supported by the nearly circular orbit. This evolutionary process would also eject the envelope of the previous AGB star, effectively halting the s-process nucleosynthesis and resulting in a diffuse nebular structure around the star. We suggest wide-angle infrared photometry or interferometry be performed to investigate this potential nebular structure. 

\paragraph{TYC 8258-1189-1} is a giant star identified by the GALAH survey as a chemically peculiar candidate. Our atmospheric parameter from Xiru are in excellent agreement with those from GALAH. This star displays a weak s-process profile in the comparison to the FRUITY yields, similar to that of the mild Ba stars HD 105671 and HD 196673, with a higher AGB mass of 5.0 M$_{\odot}$. A dilution factor of $\delta = 0.6$ suggests moderate mixing of material accreted from an AGB companion. Compared to \citet{2021MNRAS.506..150B}, we add the elements Sr, Nd, and a Pb upper limit to the pattern. We find slightly lower enhancements in Ba compared to the analysis of \citet{2021MNRAS.506..150B}, but the rest of our computed abundances are in good agreement. Our FRUITY fit suggests a higher mass AGB provided the material to produce the mild s-process signature. We propose adding this star to the mild Ba star category upon spectroscopic confirmation of the binary orbit. We do not yet detect significant RV variations in our observations, but the \emph{Gaia} DR3 RUWE parameter of $1.53$ suggests a binary system from astrometric orbital fitting. Planned follow-up RV measurements will reveal the spectroscopic nature of this orbit.

\paragraph{TYC 2250-1047-1} was identified as a Ba-rich candidate in the LAMOST data set and shown by \citet{2021A&A...654A.140K} to have strong s-process enhancement. Compared to \citet{2021A&A...654A.140K}, Xiru finds a higher temperature and a slightly higher metallicity, likely a result of the ionization and excitation balance. Our analysis confirms the heavy metal enrichment ([s/Fe] $\sim$ 0.79), and we add the elements Mo, Nd, Eu, and Pb to the abundance pattern for this star. With our additions to the abundance pattern, we find a good fit to the FRUITY yields with an initial AGB mass of 2.0 M$_{\odot}$  with significant mixing. Our Eu abundances is in good agreement with the FRUITY models in our fit, although \cite{2023A&A...677A..47K} suggest enrichment could come from the i-process. \citet{2021A&A...654A.140K} hint at the binary of the system from a variable RV; however with only 4 RV data points we cannot confidently fit an RV curve to the data. With further observations planned in our RV monitoring program, we will have sufficient data to characterize the binary orbit. 

\paragraph{HD 51273} is a poorly studied high-proper motion G-type main sequence dwarf. The stellar parameters from Xiru are in good agreement with those from \emph{Gaia}, although we estimate a higher metallicity of [Fe/H] = -0.31 $\pm$ 0.12 compared to -0.79 from \citet{2023A&A...674A...1G}. We detect very mild enhancements s-process elements ([s/Fe] = 0.06), but within uncertainties this is not significantly high compared to solar. We add 10 elements (C, Mg, Sr, Y, Zr, Mo, Ba, La, Ce, and Nd) to the abundance pattern. Comparing to the FRUITY yields, we find our observations are best fit by a 6.0 $M_{\odot}$ AGB star with little mixing ($\delta = 0.7$). In the known mild Ba stars HD 105671 and HD 196673, we find similar model fits and similar abundance patterns as in this unclassified star HD 51273. Although similar to known mild barium stars, we find the abundance pattern alone inconclusive in categorizing this star with other mild Ba stars. While we see significant RV variation between observations of HD 51273, we do not have enough data to characterize an orbit. While a Pb detection would help solidify this conjecture, without a confirmed orbit we hesitate to categorize this star as a Ba star, when it could be lightly enriched via the ambient ISM upon its formation.

\paragraph{HD 33363} is a G or K-type sub-giant or giant star with a low metallicity, and is classified as an RS-CVn type star. Comparing our stellar parameters from Xiru to those estimated by \emph{Gaia}, we find a higher temperature with a lower surface gravity and metallicity. These differences may be due to the active nature of this star. We find some scatter in the heavy element abundances, and fits to the FRUITY database result in a purely flat abundance pattern, and the dilution factor of $\delta = 0.0$ shows that this system is not indicative of AGB mass transfer. Our orbital analysis was ultimately inconclusive in constraining the binary orbit.

\paragraph{TYC 2866-338-1} is a chemically peculiar star with large enhancements in s-process material [s/Fe]=0.70. Stellar parameters are in decent agreement with those derived using the \emph{Gaia} XP spectra, albeit with a slightly higher effective temperature and surface gravity. Our computed abundances show good agreement with yields from a 3.0 M$_{\odot}$ AGB star - this is a promising result, and we suggest this star be classified as a Ba star based on the abundance pattern. The low dilution factor $\delta = 0.3$ suggests significant mixing, but does not erase the observed strong s-process abundance pattern. Our observed RV data shows significant variations between observations, and differs from the measured \emph{Gaia} DR2 systemic velocity, hinting at binarity in the system. However, we lack sufficient data to fully fit the orbit. With our RV monitoring program, we will add more data points to characterize the binary. With the confirmed orbit, we would feel more comfortable classifying this star as a Ba giant.

\paragraph{BD-19 3868} is a spectroscopic binary confirmed by RAVE \cite{2020yCat.3283....0S} and \emph{Gaia} DR3, and we find good agreement comparing atmospheric parameters derived by Xiru. Our abundance analysis results in a lightly scattered, relatively flat pattern with our [Fe/H] and [Mg/Fe] in agreement with the analysis performed by \citet{2020AJ....160...82S}. With small heavy element enhancements where [s/Fe] = $0.14$, and a Pb detection of approximately solar, we estimate a rather high mass (M$_{AGB} = 6.0$M$_{\odot}$) with a low dilution factor, meaning what material may have been transferred from an AGB companion would be heavily mixed. This abundance pattern is similar to that of the mild Ba stars HD 105671 and HD 196673. With our few snapshot RV observations, we detect significant deviation from the systemic velocity, providing further evidence to the binarity of this star, but lack enough data points to fully constrain the orbit. Despite similar abundance patterns, we hesitate to claim that this star has been mildly enriched by a previous AGB companion.

\paragraph{HD 276679} is a metal-poor ([Fe/H] = -0.90 $\pm$ 0.18) multiple-star, with the brightest component being an F-type sub-giant star ($\log g = 3.18$). Using Xiru, we find a lower temperature, surface gravity, and metallicity compared to \cite{2020AJ....160...83S}. We find enhancements in carbon with [C/Fe]$>$0.7, and are able to establish upper limits for Zr, Mo, Nd, and Eu. We find significant enhancements in s-process elements Sr, Ba, La, and Ce. Comparing our abundance pattern to the FRUITY shows a good match with a 3.0 $M_{\odot}$ model, and the overall pattern displays the prominent double-peaked profile. With the high carbon abundance and strong s-process pattern, we suggest this star may be a CH star that has been enhanced from a previous AGB companion. There are too few RV data points to detect notable variability and characterize the necessary binary orbit to make a full conclusion about the origin of the s-process signature, but future observations will provide the necessary data for an orbital analysis.

\section{Line lists for abundances}
Here we compile the line wavelengths, excitation potentials and oscillator strengths of atomic lines used in this study to compute the elemental abundances from spectral synthesis or equivalent widths. In addition to this line list, we use Fe lines from \citet{2010A&A...513A..35A, 2016A&A...587A.124K} for metal rich stars, and \citet{2012A&A...545A..31H, 2016A&A...587A.124K} for metal poor stars with ARES to compute equivalent widths of Fe I and II features for the spectroscopic iron abundances.

\begin{table*}[]
    \centering
    \caption{Wavelengths, excitation potentials, and oscillator strengths of atomic lines used in this study, sorted by element, with NIST data quality flags.}
    \begin{tabular}{| l c c l | l c c l |}
    \hline \hline
        Element / $\lambda$ & Exc. Potential & Osc. Str. & NIST Flag & Element / $\lambda$ & Exc. Potential & Osc. Str. & NIST Flag \\ 
        \AA & eV & & & \AA & eV & & \\
        \hline \hline
        Mg (12)       &       &        &         & Fe (26)       &       &        &    \\
        5528.405 (I)  & 4.343 & -0.620 & B+      & 5191.455 (I)  & 3.038 & -0.551 & B  \\
        5711.088 (I)  & 4.343 & -1.830 & B+      & 5194.942 (I)  & 1.557 & -2.090 & A  \\
                      &       &        &         & 5198.711 (I)  & 2.223 & -2.135 & B  \\
        Sr (38)       &       &        &         & 5215.179 (I)  & 3.266 & -0.871 & B  \\
        4077.714 (II) & 0.000 &  0.150 & AA      & 5229.845 (I)  & 3.283 & -0.967 &    \\
        4215.524 (II) & 0.000 & -0.170 & AA      & 5232.940 (I)  & 2.940 & -0.058 & B+ \\
        4607.331 (I)  & 0.000 &  0.280 & AA      & 5242.491 (I)  & 3.634 & -0.967 & B  \\
        4722.278 (I)  & 1.797 & -0.130 & B+      & 5247.050 (I)  & 0.087 & -4.946 & A  \\
        4962.263 (I)  & 1.846 &  0.250 & AA      & 5169.033 (II) & 2.891 & -1.250 & C  \\
        5256.899 (I)  & 2.270 &  0.390 & B+      & 5197.577 (II) & 3.230 & -2.348 & C  \\
                     &       &         &         & 5234.625 (II) & 3.221 & -2.279 &    \\
        Zr (40)      &       &         &         &               &       &        &    \\
        4359.74 (II) & 1.236 & -0.5100 &         & Y (39)        &       &        &    \\     
        4496.97 (II) & 0.713 & -0.8900 &         & 4374.933 (II) & 0.408 &  0.160 & A  \\     
        4687.80 (I)  & 0.730 &  0.5500 &         & 4689.767 (I)  & 2.002 & -0.170 & A  \\     
        4739.48 (I)  & 0.650 &  0.2300 &         & 4883.682 (II) & 1.083 &  0.070 & B  \\     
        6127.44 (I)  & 0.154 & -1.0600 &         & 5205.722 (II) & 1.032 & -0.340 & B+ \\     
                     &       &         &         & 5503.464 (I)  & 1.965 &  0.370 & A  \\     
        Ba (56)       &       &        &         & 5509.894 (II) & 0.992 & -1.010 & B+ \\     
        4554.033 (II) & 0.000 & -0.562 & B (hfs) &               &       &        &    \\
        5853.675 (II) & 0.604 & -1.914 & B (hfs) & Mo (42)       &       &        &    \\
        6141.713 (II) & 0.704 & -1.158 & B (hfs) & 5506.493 (I)  & 1.334 &  0.060 & A+ \\
        6496.898 (II) & 0.604 & -1.111 & B (hfs) & 5533.031 (I)  & 1.334 & -0.070 & A+ \\
                      &       &        &         & 5570.444 (I)  & 1.334 & -0.340 & A  \\
        Ce (58)       &       &        &         &               &       &        &    \\ 
        4628.161 (II) & 0.516 &  0.140 & B+      & La (57)       &       &        &    \\
        5274.229 (II) & 1.044 &  0.130 & B+      & 4920.98 (II) & 0.126 & -0.580 & B+       \\
        5330.556 (II) & 0.869 & -0.400 & B       & 4921.79 (II) & 0.244 & -0.450 & B+       \\
        5353.524 (II) & 0.879 &  0.090 & B+      & 5114.56 (II) & 0.235 & -1.999 & B+ (hfs) \\
                      &       &        &         & 5122.99 (II) & 0.321 & -2.212 & B+ (hfs) \\
        Eu (63)      &       &       &           & 5301.98 (II) & 0.403 & -2.300 & B+ (hfs) \\
        4129.70 (II) & 0.000 &  0.109 & (hfs)    & 5805.78 (II) & 0.126 & -2.940 & B$\;\;\;$(hfs) \\
        4594.03 (I)  & 0.000 &  0.680 & B+       &               &       &        &    \\
        5577.14 (I)  & 1.667 &  0.000 & B+       & Nd (60)       &       &        &    \\
        6645.11 (II) & 1.379 & -0.517 & (hfs)    & 4232.378 (II) &  - -  &   - -  & B  \\
                      &       &        &         & 4247.367 (II) & 0.000 & -0.210 & B+ \\
        Pb (82)      &       &        &          & 5212.365 (II) & 0.204 & -0.960 & B+ \\
        4057.807 (I) & 1.319 & -0.170 & C+       & 5361.474 (II) & 0.680 & -0.370 &    \\
        \hline     \end{tabular}
    \tablefoot{Oscillator strengths for lines that include hyper-fine splitting (hfs) are the total $\log(gf)$ from all contributing components. The ionization state is identified by the roman numerals in the parentheses, where (I) is the ground state, and (II) is the first ionized state.}
    \label{tab:line_list}
\end{table*}

\FloatBarrier

\section{Abundance uncertainties}

The average uncertainty in our abundance measurements is a combination of the statistical scatter measurement between lines and the uncertainties in the atmospheric parameters. Abundance uncertainties are calculated for all elements using the methodology described in \citet{2018A&A...618A..32K, 2021A&A...654A.140K}, following Equation (2) from \citet{2002ApJS..139..219J} or Equation (A20) from \citet{1995AJ....109.2757M}, where $\sigma_{ran}$ is the statistical uncertainty in the line measurements, and $\sigma_T$ , $\sigma_{\log g}$, $\sigma_{[Fe/H]}$, and $\sigma_{\xi}$ are the uncertainties on the atmospheric parameters:

\begin{align*}\label{eqn:uncertainties}
    \sigma_{tot}^2 = \sigma_{ran}^2 + \left(\frac{\partial \log\epsilon}{\partial T}\right)^2\sigma_{T_eff}^2 + \left(\frac{\partial \log\epsilon}{\partial\log g}\right)^2\sigma_{\log g}^2 + \\ 
    \left(\frac{\partial\log\epsilon}{\partial\xi}\right)^2\sigma_{\xi}^2 + \left(\frac{\partial\log\epsilon}{\partial [Fe/H]}\right)^2\sigma_{[Fe/H]}^2 + \\
    2 \Bigg( \left(\frac{\partial\log\epsilon}{\partial T}\right)\left(\frac{\partial\log\epsilon}{\partial\log g}\right)\sigma_{T,\log g} + \left(\frac{\partial\log\epsilon}{\partial\log g}\right)\left(\frac{\partial\log\epsilon}{\partial\xi}\right)\sigma_{\log g,\xi} + \\
    \left(\frac{\partial\log\epsilon}{\partial\xi}\right)\left(\frac{\partial\log\epsilon}{\partial T}\right)\sigma_{\xi,T} + \left(\frac{\partial\log\epsilon}{\partial\ T}\right)\left(\frac{\partial\log\epsilon}{\partial [Fe/H]}\right)\sigma_{\log g,\xi} + \\
    \left(\frac{\partial\log\epsilon}{\partial\log g}\right)\left(\frac{\partial\log\epsilon}{\partial [Fe/H]}\right)\sigma_{\log g,\xi} + \left(\frac{\partial\log\epsilon}{\partial\xi}\right)\left(\frac{\partial\log\epsilon}{\partial [Fe/H]}\right)\sigma_{\log g,\xi} \Bigg) .
\end{align*}

\section{Correlations in abundance space} 

\begin{figure*}
    \centering
    \includegraphics[width=\linewidth]{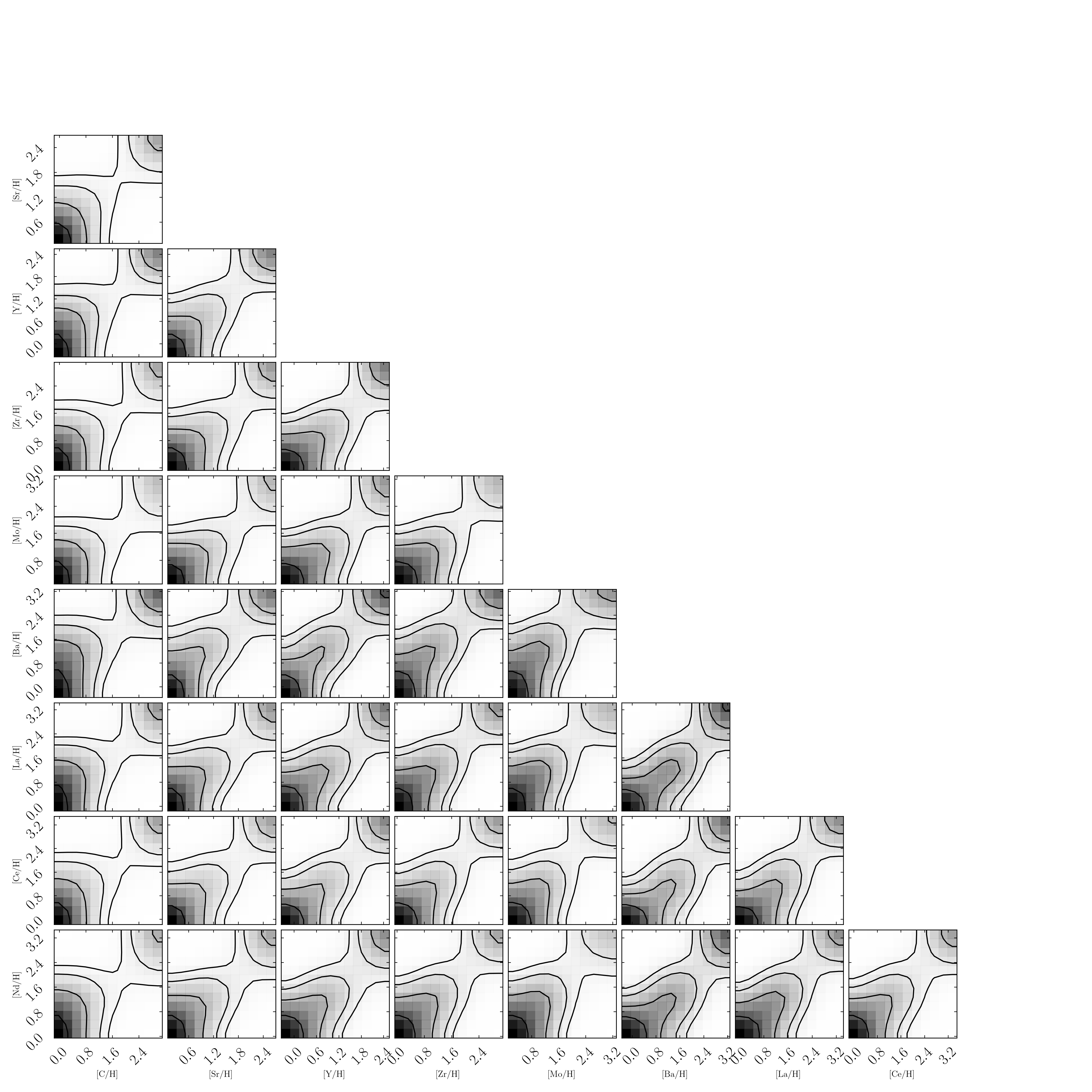}
    \caption{Corner plot of s-process materials. We find correlations between s-process elements within the light and heavy peaks respectively, as well as between light and heavy s-elements. There are strong correlations between heavy s-elements and Eu.}
    \label{fig:abundance_corner}
\end{figure*}

\end{appendix}

\end{document}